\documentstyle[12pt,a4,epsfig]{article}
\newcommand{\neutr}{\chi}
\newcommand{\sneut}{\tilde{\nu}}

\newcommand{\lesssim}{\,\raisebox{-0.6ex}{$\stackrel{\textstyle<}{\textstyle\sim}$}\,}

\def \stop {\tilde{\mathrm{t}}}
\def \sbot {\tilde{\mathrm{b}}}
\def \snu {\tilde{\nu}}
\def \neu {\chi}
\def \deltm {\Delta m}
\def \thetamix  {\theta_{\mathrm{\tilde{t}}}}
\def \thetab  {\theta_{\mathrm{\tilde{b}}}}
\def \gg   {\gamma\gamma}
\def \ggqq {\gamma\gamma \rightarrow q\bar{q}}
\def \ggtt {\mathrm \gamma\gamma \rightarrow \tau^{+}\tau^{-}}

\def \qqg  {q\bar{q}(\gamma)}

\def \ww   {\mathrm WW}
\def \zz   {\mathrm Z\gamma^{*}}

\def \zee  {\mathrm Zee}
\def \ewnu {\mathrm We\nu}
\def \rts  {\sqrt{s}}
\def \pt   {p_{\mathrm{t}}}
\def \diffa {\theta_{\mathrm{point}}}
\def \thscat {\theta_{\mathrm{scat}}}
\def \mstop {m_{\tilde{\mathrm{t}}}}

\def \elow   {E_{12^{\circ}}}
\def \stopr {\mathrm \tilde{t}_{R}}
\def \stopl {\mathrm \tilde{t}_{L}}
\def \thmiss {\theta_{P_{\mathrm{miss}}}}
\def \gev  { \, \mathrm{GeV}/\it{c}^{\mathrm{2}}}
\def \gvm  { \, \mathrm{GeV}/\it{c}}
\def \mx   {M_{\mathrm{eff}}} 
\def \nch  {N_{\mathrm{ch}}}

\begin{document}
\pagestyle{empty}
\begin{center} EUROPEAN LABORATORY FOR PARTICLE PHYSICS (CERN) \end{center}
\bigskip
\begin{flushright} 
CERN-PPE/97-084\\
17 July 1997
\vspace{1cm}
\end{flushright} 
\vspace{2cm}
\begin{center}
  \mathversion{bold}
  {\LARGE\bf Searches for Scalar Top and Scalar Bottom Quarks 
              at LEP2}
  \mathversion{normal}
  \vskip 1.2cm
  {\bf The ALEPH Collaboration}\\
  \vskip 1.2cm
  {\bf Abstract}
\end{center}
\medskip
%
%
%
\small

Searches for scalar top and bottom quarks have been performed with
data collected by the ALEPH detector at LEP.  The data sample consists
of 21.7~$\mathrm{pb}^{-1}$ taken at $\rts$ = 161, 170, and 172~GeV and  
5.7~$\mathrm{pb}^{-1}$ taken at $\rts$ = 130 and 136~GeV.  
No evidence for scalar top quarks or scalar bottom quarks was found 
in the channels $\stop \rightarrow \mathrm{c}\neu$, 
$\stop \rightarrow \mathrm{b}\ell\snu$, and $\sbot \rightarrow \mathrm{b}\neu$.
For the channel $\stop \rightarrow \mathrm{c}\neu$ a limit of 
67~$\gev$ has been set on the scalar top quark mass, independent of the 
mixing angle between the supersymmetric partners of the left and right-handed
states of the top quark. This limit assumes a mass difference between the
$\stop$ and the $\neu$ of at least 10~$\gev$. For the channel
$\stop \rightarrow \mathrm{b}\ell\snu$ the mixing-angle independent scalar top
limit is 70~$\gev$, assuming a mass difference between the $\stop$ and the
$\snu$ of at least 10~$\gev$. 
For the channel $\sbot \rightarrow \mathrm{b}\neu$, a limit of 73~$\gev$ 
has been set on the mass of the supersymmetric partner of the
left-handed state of the bottom quark. This limit is valid 
if the mass difference between the $\sbot$ and the $\chi$ is 
at least 10~$\gev$.


\vspace{1cm}
\begin{center}
\vspace{1cm}
{\em (To be submitted to Physics Letters B)}
\end{center}

\vfill
\pagebreak
%
%
\pagestyle{empty}
\newpage
\small
%
%
\newlength{\saveparskip}
\newlength{\savetextheight}
\newlength{\savetopmargin}
\newlength{\savetextwidth}
\newlength{\saveoddsidemargin}
\newlength{\savetopsep}
\setlength{\saveparskip}{\parskip}
\setlength{\savetextheight}{\textheight}
\setlength{\savetopmargin}{\topmargin}
\setlength{\savetextwidth}{\textwidth}
\setlength{\saveoddsidemargin}{\oddsidemargin}
\setlength{\savetopsep}{\topsep}
%
%
\setlength{\parskip}{0.0cm}
\setlength{\textheight}{25.0cm}
\setlength{\topmargin}{-1.5cm}
\setlength{\textwidth}{16 cm}
\setlength{\oddsidemargin}{-0.0cm}
\setlength{\topsep}{1mm}
\pretolerance=10000
\centerline{\large\bf The ALEPH Collaboration}
\footnotesize
\vspace{0.5cm}
{\raggedbottom
\begin{sloppypar}
\samepage\noindent
R.~Barate,
D.~Buskulic,
D.~Decamp,
P.~Ghez,
C.~Goy,
J.-P.~Lees,
A.~Lucotte,
M.-N.~Minard,
J.-Y.~Nief,
B.~Pietrzyk
\nopagebreak
\begin{center}
\parbox{15.5cm}{\sl\samepage
Laboratoire de Physique des Particules (LAPP), IN$^{2}$P$^{3}$-CNRS,
74019 Annecy-le-Vieux Cedex, France}
\end{center}\end{sloppypar}
\vspace{2mm}
\begin{sloppypar}
\noindent
M.P.~Casado,
M.~Chmeissani,
P.~Comas,
J.M.~Crespo,
M.~Delfino, 
E.~Fernandez,
M.~Fernandez-Bosman,
Ll.~Garrido,$^{15}$
A.~Juste,
M.~Martinez,
G.~Merino,
R.~Miquel,
Ll.M.~Mir,
C.~Padilla,
I.C.~Park,
A.~Pascual,
J.A.~Perlas,
I.~Riu,
F.~Sanchez,
F.~Teubert
\nopagebreak
\begin{center}
\parbox{15.5cm}{\sl\samepage
Institut de F\'{i}sica d'Altes Energies, Universitat Aut\`{o}noma
de Barcelona, 08193 Bellaterra (Barcelona), Spain$^{7}$}
\end{center}\end{sloppypar}
\vspace{2mm}
\begin{sloppypar}
\noindent
A.~Colaleo,
D.~Creanza,
M.~de~Palma,
G.~Gelao,
G.~Iaselli,
G.~Maggi,
M.~Maggi,
N.~Marinelli,
S.~Nuzzo,
A.~Ranieri,
G.~Raso,
F.~Ruggieri,
G.~Selvaggi,
L.~Silvestris,
P.~Tempesta,
A.~Tricomi,$^{3}$
G.~Zito
\nopagebreak
\begin{center}
\parbox{15.5cm}{\sl\samepage
Dipartimento di Fisica, INFN Sezione di Bari, 70126
Bari, Italy}
\end{center}\end{sloppypar}
\vspace{2mm}
\begin{sloppypar}
\noindent
X.~Huang,
J.~Lin,
Q. Ouyang,
T.~Wang,
Y.~Xie,
R.~Xu,
S.~Xue,
J.~Zhang,
L.~Zhang,
W.~Zhao
\nopagebreak
\begin{center}
\parbox{15.5cm}{\sl\samepage
Institute of High-Energy Physics, Academia Sinica, Beijing, The People's
Republic of China$^{8}$}
\end{center}\end{sloppypar}
\vspace{2mm}
\begin{sloppypar}
\noindent
D.~Abbaneo,
R.~Alemany,
A.O.~Bazarko,$^{1}$
U.~Becker,
P.~Bright-Thomas,
M.~Cattaneo,
F.~Cerutti,
G.~Dissertori,
H.~Drevermann,
R.W.~Forty,
M.~Frank,
R.~Hagelberg,
J.B.~Hansen,
J.~Harvey,
P.~Janot,
B.~Jost,
E.~Kneringer,
J.~Knobloch,
I.~Lehraus,
P.~Mato,
A.~Minten,
L.~Moneta,
A.~Pacheco,
J.-F.~Pusztaszeri,$^{20}$
F.~Ranjard,
G.~Rizzo,
L.~Rolandi,
D.~Rousseau,
D.~Schlatter,
M.~Schmitt,
O.~Schneider,
W.~Tejessy,
I.R.~Tomalin,
H.~Wachsmuth,
A.~Wagner$^{21}$
\nopagebreak
\begin{center}
\parbox{15.5cm}{\sl\samepage
European Laboratory for Particle Physics (CERN), 1211 Geneva 23,
Switzerland}
\end{center}\end{sloppypar}
\vspace{2mm}
\begin{sloppypar}
\noindent
Z.~Ajaltouni,
A.~Barr\`{e}s,
C.~Boyer,
A.~Falvard,
C.~Ferdi,
P.~Gay,
C~.~Guicheney,
P.~Henrard,
J.~Jousset,
B.~Michel,
S.~Monteil,
J-C.~Montret,
D.~Pallin,
P.~Perret,
F.~Podlyski,
J.~Proriol,
P.~Rosnet,
J.-M.~Rossignol
\nopagebreak
\begin{center}
\parbox{15.5cm}{\sl\samepage
Laboratoire de Physique Corpusculaire, Universit\'e Blaise Pascal,
IN$^{2}$P$^{3}$-CNRS, Clermont-Ferrand, 63177 Aubi\`{e}re, France}
\end{center}\end{sloppypar}
\vspace{2mm}
\begin{sloppypar}
\noindent
T.~Fearnley,
J.D.~Hansen,
J.R.~Hansen,
P.H.~Hansen,
B.S.~Nilsson,
B.~Rensch,
A.~W\"a\"an\"anen
\begin{center}
\parbox{15.5cm}{\sl\samepage
Niels Bohr Institute, 2100 Copenhagen, Denmark$^{9}$}
\end{center}\end{sloppypar}
\vspace{2mm}
\begin{sloppypar}
\noindent
G.~Daskalakis,
A.~Kyriakis,
C.~Markou,
E.~Simopoulou,
A.~Vayaki
\nopagebreak
\begin{center}
\parbox{15.5cm}{\sl\samepage
Nuclear Research Center Demokritos (NRCD), Athens, Greece}
\end{center}\end{sloppypar}
\vspace{2mm}
\begin{sloppypar}
\noindent
A.~Blondel,
J.C.~Brient,
F.~Machefert,
A.~Roug\'{e},
M.~Rumpf,
A.~Valassi,$^{6}$
H.~Videau
\nopagebreak
\begin{center}
\parbox{15.5cm}{\sl\samepage
Laboratoire de Physique Nucl\'eaire et des Hautes Energies, Ecole
Polytechnique, IN$^{2}$P$^{3}$-CNRS, 91128 Palaiseau Cedex, France}
\end{center}\end{sloppypar}
\vspace{2mm}
\begin{sloppypar}
\noindent
E.~Focardi,
G.~Parrini,
K.~Zachariadou
\nopagebreak
\begin{center}
\parbox{15.5cm}{\sl\samepage
Dipartimento di Fisica, Universit\`a di Firenze, INFN Sezione di Firenze,
50125 Firenze, Italy}
\end{center}\end{sloppypar}
\vspace{2mm}
\begin{sloppypar}
\noindent
R.~Cavanaugh,
M.~Corden,
C.~Georgiopoulos,
T.~Huehn,
D.E.~Jaffe
\nopagebreak
\begin{center}
\parbox{15.5cm}{\sl\samepage
Supercomputer Computations Research Institute,
Florida State University,
Tallahassee, FL 32306-4052, USA $^{13,14}$}
\end{center}\end{sloppypar}
\vspace{2mm}
\begin{sloppypar}
\noindent
A.~Antonelli,
G.~Bencivenni,
G.~Bologna,$^{4}$
F.~Bossi,
P.~Campana,
G.~Capon,
D.~Casper,
V.~Chiarella,
G.~Felici,
P.~Laurelli,
G.~Mannocchi,$^{5}$
F.~Murtas,
G.P.~Murtas,
L.~Passalacqua,
M.~Pepe-Altarelli
\nopagebreak
\begin{center}
\parbox{15.5cm}{\sl\samepage
Laboratori Nazionali dell'INFN (LNF-INFN), 00044 Frascati, Italy}
\end{center}\end{sloppypar}
\vspace{2mm}
\begin{sloppypar}
\noindent
L.~Curtis,
S.J.~Dorris,
A.W.~Halley,
I.G.~Knowles,
J.G.~Lynch,
V.~O'Shea,
C.~Raine,
J.M.~Scarr,
K.~Smith,
P.~Teixeira-Dias,
A.S.~Thompson,
E.~Thomson,
F.~Thomson,
R.M.~Turnbull
\nopagebreak
\begin{center}
\parbox{15.5cm}{\sl\samepage
Department of Physics and Astronomy, University of Glasgow, Glasgow G12
8QQ,United Kingdom$^{10}$}
\end{center}\end{sloppypar}
\vspace{2mm}
\begin{sloppypar}
\noindent
O.~Buchm\"uller,
S.~Dhamotharan,
C.~Geweniger,
G.~Graefe,
P.~Hanke,
G.~Hansper,
V.~Hepp,
E.E.~Kluge,
A.~Putzer,
J.~Sommer,
K.~Tittel,
S.~Werner,
M.~Wunsch
\begin{center}
\parbox{15.5cm}{\sl\samepage
Institut f\"ur Hochenergiephysik, Universit\"at Heidelberg, 69120
Heidelberg, Fed.\ Rep.\ of Germany$^{16}$}
\end{center}\end{sloppypar}
\vspace{2mm}
\begin{sloppypar}
\noindent
R.~Beuselinck,
D.M.~Binnie,
W.~Cameron,
P.J.~Dornan,
M.~Girone,
S.~Goodsir,
E.B.~Martin,
P.~Morawitz,
A.~Moutoussi,
J.~Nash,
J.K.~Sedgbeer,
P.~Spagnolo,
A.M.~Stacey,
M.D.~Williams
\nopagebreak
\begin{center}
\parbox{15.5cm}{\sl\samepage
Department of Physics, Imperial College, London SW7 2BZ,
United Kingdom$^{10}$}
\end{center}\end{sloppypar}
\vspace{2mm}
\begin{sloppypar}
\noindent
V.M.~Ghete,
P.~Girtler,
D.~Kuhn,
G.~Rudolph
\nopagebreak
\begin{center}
\parbox{15.5cm}{\sl\samepage
Institut f\"ur Experimentalphysik, Universit\"at Innsbruck, 6020
Innsbruck, Austria$^{18}$}
\end{center}\end{sloppypar}
\vspace{2mm}
\begin{sloppypar}
\noindent
A.P.~Betteridge,
C.K.~Bowdery,
P.~Colrain,
G.~Crawford,
A.J.~Finch,
F.~Foster,
G.~Hughes,
R.W.L.~Jones,
T.~Sloan,
E.P.~Whelan,
M.I.~Williams
\nopagebreak
\begin{center}
\parbox{15.5cm}{\sl\samepage
Department of Physics, University of Lancaster, Lancaster LA1 4YB,
United Kingdom$^{10}$}
\end{center}\end{sloppypar}
\vspace{2mm}
\begin{sloppypar}
\noindent
C.~Hoffmann,
K.~Jakobs,
K.~Kleinknecht,
G.~Quast,
B.~Renk,
E.~Rohne,
H.-G.~Sander,
P.~van~Gemmeren,
C.~Zeitnitz
\nopagebreak
\begin{center}
\parbox{15.5cm}{\sl\samepage
Institut f\"ur Physik, Universit\"at Mainz, 55099 Mainz, Fed.\ Rep.\
of Germany$^{16}$}
\end{center}\end{sloppypar}
\vspace{2mm}
\begin{sloppypar}
\noindent
J.J.~Aubert,
C.~Benchouk,
A.~Bonissent,
G.~Bujosa,
J.~Carr,
P.~Coyle,
C.~Diaconu,
A.~Ealet,
D.~Fouchez,
N.~Konstantinidis,
O.~Leroy,
F.~Motsch,
P.~Payre,
M.~Talby,
A.~Sadouki,
M.~Thulasidas,
A.~Tilquin,
K.~Trabelsi
\nopagebreak
\begin{center}
\parbox{15.5cm}{\sl\samepage
Centre de Physique des Particules, Facult\'e des Sciences de Luminy,
IN$^{2}$P$^{3}$-CNRS, 13288 Marseille, France}
\end{center}\end{sloppypar}
\vspace{2mm}
\begin{sloppypar}
\noindent
M.~Aleppo, 
M.~Antonelli,
F.~Ragusa$^{12}$
\nopagebreak
\begin{center}
\parbox{15.5cm}{\sl\samepage
Dipartimento di Fisica, Universit\`a di Milano e INFN Sezione di
Milano, 20133 Milano, Italy.}
\end{center}\end{sloppypar}
\vspace{2mm}
\begin{sloppypar}
\noindent
R.~Berlich,
W.~Blum,
V.~B\"uscher,
H.~Dietl,
G.~Ganis,
C.~Gotzhein,
H.~Kroha,
G.~L\"utjens,
G.~Lutz,
W.~M\"anner,
H.-G.~Moser,
R.~Richter,
A.~Rosado-Schlosser,
S.~Schael,
R.~Settles,
H.~Seywerd,
R.~St.~Denis,
H.~Stenzel,
W.~Wiedenmann,
G.~Wolf
\nopagebreak
\begin{center}
\parbox{15.5cm}{\sl\samepage
Max-Planck-Institut f\"ur Physik, Werner-Heisenberg-Institut,
80805 M\"unchen, Fed.\ Rep.\ of Germany\footnotemark[16]}
\end{center}\end{sloppypar}
\vspace{2mm}
\begin{sloppypar}
\noindent
J.~Boucrot,
O.~Callot,$^{12}$
S.~Chen,
A.~Cordier,
M.~Davier,
L.~Duflot,
J.-F.~Grivaz,
Ph.~Heusse,
A.~H\"ocker,
A.~Jacholkowska,
M.~Jacquet,
D.W.~Kim,$^{2}$
F.~Le~Diberder,
J.~Lefran\c{c}ois,
A.-M.~Lutz,
I.~Nikolic,
M.-H.~Schune,
L.~Serin,
S.~Simion,
E.~Tournefier,
J.-J.~Veillet,
I.~Videau,
D.~Zerwas
\nopagebreak
\begin{center}
\parbox{15.5cm}{\sl\samepage
Laboratoire de l'Acc\'el\'erateur Lin\'eaire, Universit\'e de Paris-Sud,
IN$^{2}$P$^{3}$-CNRS, 91405 Orsay Cedex, France}
\end{center}\end{sloppypar}
\vspace{2mm}
\begin{sloppypar}
\noindent
\samepage
P.~Azzurri,
G.~Bagliesi,
S.~Bettarini,
C.~Bozzi,
G.~Calderini,
V.~Ciulli,
R.~Dell'Orso,
R.~Fantechi,
I.~Ferrante,
A.~Giassi,
A.~Gregorio,
F.~Ligabue,
A.~Lusiani,
P.S.~Marrocchesi,
A.~Messineo,
F.~Palla,
G.~Sanguinetti,
A.~Sciab\`a,
G.~Sguazzoni,
J.~Steinberger,
R.~Tenchini,
C.~Vannini,
A.~Venturi,
P.G.~Verdini
\samepage
\begin{center}
\parbox{15.5cm}{\sl\samepage
Dipartimento di Fisica dell'Universit\`a, INFN Sezione di Pisa,
e Scuola Normale Superiore, 56010 Pisa, Italy}
\end{center}\end{sloppypar}
\vspace{2mm}
\begin{sloppypar}
\noindent
G.A.~Blair,
L.M.~Bryant,
J.T.~Chambers,
Y.~Gao,
M.G.~Green,
T.~Medcalf,
P.~Perrodo,
J.A.~Strong,
J.H.~von~Wimmersperg-Toeller
\nopagebreak
\begin{center}
\parbox{15.5cm}{\sl\samepage
Department of Physics, Royal Holloway \& Bedford New College,
University of London, Surrey TW20 OEX, United Kingdom$^{10}$}
\end{center}\end{sloppypar}
\vspace{2mm}
\begin{sloppypar}
\noindent
D.R.~Botterill,
R.W.~Clifft,
T.R.~Edgecock,
S.~Haywood,
P.~Maley,
P.R.~Norton,
J.C.~Thompson,
A.E.~Wright
\nopagebreak
\begin{center}
\parbox{15.5cm}{\sl\samepage
Particle Physics Dept., Rutherford Appleton Laboratory,
Chilton, Didcot, Oxon OX11 OQX, United Kingdom$^{10}$}
\end{center}\end{sloppypar}
\vspace{2mm}
\begin{sloppypar}
\noindent
B.~Bloch-Devaux,
P.~Colas,
B.~Fabbro,
W.~Kozanecki,
E.~Lan\c{c}on,
M.C.~Lemaire,
E.~Locci,
P.~Perez,
J.~Rander,
J.-F.~Renardy,
A.~Rosowsky,
A.~Roussarie,
J.-P.~Schuller,
J.~Schwindling,
A.~Trabelsi,
B.~Vallage
\nopagebreak
\begin{center}
\parbox{15.5cm}{\sl\samepage
CEA, DAPNIA/Service de Physique des Particules,
CE-Saclay, 91191 Gif-sur-Yvette Cedex, France$^{17}$}
\end{center}\end{sloppypar}
\vspace{2mm}
\begin{sloppypar}
\noindent
S.N.~Black,
J.H.~Dann,
H.Y.~Kim,
A.M.~Litke,
M.A. McNeil,
G.~Taylor
\nopagebreak
\begin{center}
\parbox{15.5cm}{\sl\samepage
Institute for Particle Physics, University of California at
Santa Cruz, Santa Cruz, CA 95064, USA$^{19}$}
\end{center}\end{sloppypar}
\pagebreak
\vspace{2mm}
\begin{sloppypar}
\noindent
C.N.~Booth,
R.~Boswell,
C.A.J.~Brew,
S.~Cartwright,
F.~Combley,
M.S.~Kelly,
M.~Lehto,
W.M.~Newton,
J.~Reeve,
L.F.~Thompson
\nopagebreak
\begin{center}
\parbox{15.5cm}{\sl\samepage
Department of Physics, University of Sheffield, Sheffield S3 7RH,
United Kingdom$^{10}$}
\end{center}\end{sloppypar}
\vspace{2mm}
\begin{sloppypar}
\noindent
K.~Affholderbach,
A.~B\"ohrer,
S.~Brandt,
G.~Cowan,
J.~Foss,
C.~Grupen,
G.~Lutters,
P.~Saraiva,
L.~Smolik,
F.~Stephan 
\nopagebreak
\begin{center}
\parbox{15.5cm}{\sl\samepage
Fachbereich Physik, Universit\"at Siegen, 57068 Siegen,
 Fed.\ Rep.\ of Germany$^{16}$}
\end{center}\end{sloppypar}
\vspace{2mm}
\begin{sloppypar}
\noindent
M.~Apollonio,
L.~Bosisio,
R.~Della~Marina,
G.~Giannini,
B.~Gobbo,
G.~Musolino
\nopagebreak
\begin{center}
\parbox{15.5cm}{\sl\samepage
Dipartimento di Fisica, Universit\`a di Trieste e INFN Sezione di Trieste,
34127 Trieste, Italy}
\end{center}\end{sloppypar}
\vspace{2mm}
\begin{sloppypar}
\noindent
J.~Putz,
J.~Rothberg,
S.~Wasserbaech,
R.W.~Williams
\nopagebreak
\begin{center}
\parbox{15.5cm}{\sl\samepage
Experimental Elementary Particle Physics, University of Washington, WA 98195
Seattle, U.S.A.}
\end{center}\end{sloppypar}
\vspace{2mm}
\begin{sloppypar}
\noindent
S.R.~Armstrong,
E.~Charles,
P.~Elmer,
D.P.S.~Ferguson,
S.~Gonz\'{a}lez,
T.C.~Greening,
O.J.~Hayes,
H.~Hu,
S.~Jin,
P.A.~McNamara III,
J.M.~Nachtman,
J.~Nielsen,
W.~Orejudos,
Y.B.~Pan,
Y.~Saadi,
I.J.~Scott,
J.~Walsh,
Sau~Lan~Wu,
X.~Wu,
J.M.~Yamartino,
G.~Zobernig
\nopagebreak
\begin{center}
\parbox{15.5cm}{\sl\samepage
Department of Physics, University of Wisconsin, Madison, WI 53706,
USA$^{11}$}
\end{center}\end{sloppypar}
}
\footnotetext[1]{Now at Princeton University, Princeton, NJ 08544, U.S.A.}
\footnotetext[2]{Permanent address: Kangnung National University, Kangnung,
Korea.}
\footnotetext[3]{Also at Dipartimento di Fisica, INFN Sezione di Catania,
Catania, Italy.}
\footnotetext[4]{Also Istituto di Fisica Generale, Universit\`{a} di
Torino, Torino, Italy.}
\footnotetext[5]{Also Istituto di Cosmo-Geofisica del C.N.R., Torino,
Italy.}
\footnotetext[6]{Supported by the Commission of the European Communities,
contract ERBCHBICT941234.}
\footnotetext[7]{Supported by CICYT, Spain.}
\footnotetext[8]{Supported by the National Science Foundation of China.}
\footnotetext[9]{Supported by the Danish Natural Science Research Council.}
\footnotetext[10]{Supported by the UK Particle Physics and Astronomy Research
Council.}
\footnotetext[11]{Supported by the US Department of Energy, grant
DE-FG0295-ER40896.}
\footnotetext[12]{Also at CERN, 1211 Geneva 23,Switzerland.}
\footnotetext[13]{Supported by the US Department of Energy, contract
DE-FG05-92ER40742.}
\footnotetext[14]{Supported by the US Department of Energy, contract
DE-FC05-85ER250000.}
\footnotetext[15]{Permanent address: Universitat de Barcelona, 08208 Barcelona,
Spain.}
\footnotetext[16]{Supported by the Bundesministerium f\"ur Bildung,
Wissenschaft, Forschung und Technologie, Fed. Rep. of Germany.}
\footnotetext[17]{Supported by the Direction des Sciences de la
Mati\`ere, C.E.A.}
\footnotetext[18]{Supported by Fonds zur F\"orderung der wissenschaftlichen
Forschung, Austria.}
\footnotetext[19]{Supported by the US Department of Energy,
grant DE-FG03-92ER40689.}
\footnotetext[20]{Now at School of Operations Research and Industrial
Engireering, Cornell University, Ithaca, NY 14853-3801, U.S.A.}
\footnotetext[21]{Now at Schweizerischer Bankverein, Basel, Switzerland.}
%
%
\setlength{\parskip}{\saveparskip}
\setlength{\textheight}{\savetextheight}
\setlength{\topmargin}{\savetopmargin}
\setlength{\textwidth}{\savetextwidth}
\setlength{\oddsidemargin}{\saveoddsidemargin}
\setlength{\topsep}{\savetopsep}
\normalsize
\newpage
\pagestyle{plain}
\setcounter{page}{1}
\section{Introduction}

In the Minimal Supersymmetric Extension of the Standard Model (MSSM)~\cite{SUSY} 
each Standard Model fermion has two scalar supersymmetric partners, 
one for each chirality state.
The scalar-tops (stops) $\stopr$ and $\stopl$ are the supersymmetric partners 
of the top quark. These two fields are weak interaction eigenstates  
which mix to form the mass eigenstates. 
The stop mass matrix is given by~\cite{Hikasa}:
$$
\left(
\begin{array}{cc}
m_{\stopl}^{2}
&  m_{\mathrm{t}}a_{\mathrm{t}} \\
m_{\mathrm{t}}a_{\mathrm{t}}  
&
m_{\stopr}^{2}
\end{array}
\right),
$$
where $m_{\stopr}$ and $m_{\stopl}$ are the  $\stopr$ and $\stopl$ mass terms, $a_{t}$ 
is related to  the soft SUSY-breaking parameter $A_t$ by
$a_t = A_t - \mu / \tan{\beta}$ (where $\mu$ is the supersymmetric mass term which mixes the two 
Higgs superfields and $\tan{\beta}$ is the ratio between their vacuum expectation values) 
and $m_{\mathrm{t}}$ is the top quark mass.
Since the off-diagonal terms of this matrix are
proportional to $ m_{\mathrm{t}}$, the mixing between the weak
interaction eigenstates may be large and the lighter stop could be
the lightest supersymmetric charged particle.
The stop mass eigenstates are obtained by a unitary transformation of the 
$\stopr$ and $\stopl$ fields, parametrised by the mixing 
angle $\thetamix$. 
The lighter stop 
is given by $\stop = \mathrm{\tilde{t}_L \cos{\thetamix}}
+ \mathrm{\tilde{t}_R \sin{\thetamix}}$, while  
the heavier stop is the orthogonal combination. 


The stop could be produced 
at LEP in pairs, $\rm{e^+ e^-} \to \stop \bar{\stop}$, via s-channel exchange of a virtual 
photon or a Z.
The production cross section~\cite{Drees} depends on the stop 
charge for the coupling to the photon and on the 
weak mixing angle $\theta_{\mathrm{W}}$ and the 
mixing angle $\thetamix$ for the coupling 
to the Z.
When $\thetamix$ is about $56^{\circ}$ the lightest stop decouples 
from the Z and its cross section is almost minimal. At $\rts$ = 172~GeV, 
the maximum cross section is of order 
1~pb for a $\stop$ mass of 60~$\gev$ and is reached for $\thetamix=0^{\circ}$.

\begin{figure}[tcb]
\begin{center}
\epsfig{file=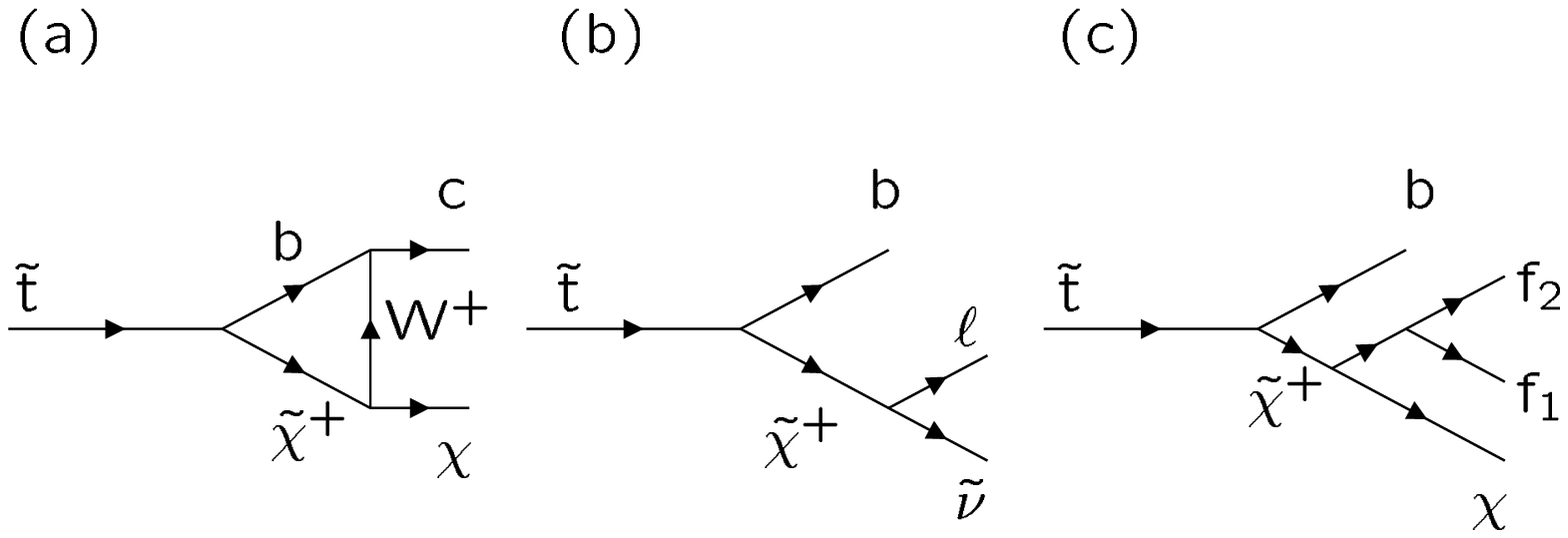,width=0.9\textwidth}
\end{center}
\caption{\rm Stop decay diagrams. (a) $\stop \rightarrow \mathrm{c}\neu$.
(b) $\stop \to \rm{b} \ell \tilde{\nu}$. 
(c) $\stop \to \mathrm{b} \mathrm{f}_{1} \mathrm{f}_{2} \neutr$. 
    Decay (c) is not
considered in this paper.
\label{stdecay}}
\end{figure}

The searches for stops described here assume that all supersymmetric
particles except the lightest neutralino $\neu$ and (possibly) the sneutrino
$\snu$ are heavier than the stop. The conservation of R-parity is also 
assumed; this implies that the Lightest Supersymmetric Particle (LSP) 
is stable. Under these assumptions,
the two dominant decay channels are $\stop \to \rm{c} \neu$ 
or $\stop \to\rm{b} \ell \tilde{\nu}$~\cite{Hikasa}.
The corresponding diagrams are shown in Figures~\ref{stdecay}a
and~\ref{stdecay}b.
The first decay can only proceed via loops and thus 
has a very small width, of the order of 1--0.01~eV~\cite{Hikasa}. 

The $\stop \to \rm{b} \ell \tilde{\nu}$ 
channel proceeds via a virtual chargino exchange 
and has a width of the order of 0.1--10~keV~\cite{Hikasa}, 
where the largest width is reached for a chargino mass close to the stop mass. 
This decay dominates when it is kinematically allowed. 
Assuming equal mass sneutrinos $\snu_{\mathrm{e}}$, 
$\snu_{\mathrm{\mu}}$ and $\snu_{\mathrm{\tau}}$, the lepton flavour for this
decay is determined by the chargino composition.  If the chargino is
the supersymmetric partner of the W the decays 
$\stop \to \mathrm{b} \mathrm{e} \snu_{\mathrm{e}}$,
$\stop \to \mathrm{b} \mu \snu_{\mu}$ and
$\stop \to \mathrm{b} \tau \snu_{\tau}$ occur with equal
branching fractions. If the chargino is the supersymmetric partner of
the charged Higgs the branching fraction of the 
decay $\stop \to \mathrm{b} \tau \snu_{\tau}$
is enhanced. In all of these cases,
if the neutralino is the LSP the sneutrino can decay 
into $(\neutr \nu)$ but this invisible decay does not change the 
experimental topology.

A possible third stop decay channel is the four-body decay $\stop \to 
\rm{b} {\it f_{\rm{1}} \bar{f}_{\rm{2}}} \neutr$. 
One such four-body decay of the $\stop$ is shown in Figure~\ref{stdecay}c.
The rates of four-body decays are expected to be much smaller than 
that of the decay $\stop \rightarrow \rm{c} \neu$. 


The phenomenology of the scalar bottom (sbottom), the supersymmetric 
partner of the bottom quark, is similar to the phenomenology of the
stop. 
In contrast to stops, sbottom mixing
is expected to be large for large values of $\tan\beta$, because of the 
relation $a_b = A_b - \mu \tan{\beta}$.
When the sbottom mixing angle $\thetab$ is about $68^{\circ}$ 
the lightest sbottom decouples from the Z.
Assuming that the $\sbot$ is lighter than all supersymmetric particles except
the $\neu$, the $\sbot$ will decay as $\mathrm{\tilde{b} \rightarrow
b \neu}$.  Compared to the $\stop$ decays, the $\sbot$ decay has
a large width of the order of 10--100~MeV.

Direct searches for stops and sbottoms 
are performed for the stop decay channels $\stop \to \rm{c}\neu$ and
$\stop \to\rm{b} \ell \tilde{\nu}$
and for the sbottom decay channel $\sbot \to \rm{b} \neu$. 
The results of these searches supersede the ALEPH results reported 
earlier for data collected at energies up to 
$\rts$ = 136~GeV~\cite{ALEPH_stop}.
The D0 experiment~\cite{D0} has reported a lower limit on the stop mass 
of 85~${\mathrm GeV}/c^2$ for the decay into $\rm{c} \neutr$ and for
a mass difference between the $\stop$ and the 
$\chi$ larger than about 40~$\gev$.
Searches for $\stop \rightarrow \rm{c} \neutr$, 
$\stop\rightarrow\mathrm{b} \ell \tilde{\nu}$ and
$\sbot\rightarrow\rm{b}\neutr$ using data
collected at LEP at energies up to $\sqrt{s}$ = 172~$\mathrm{GeV}$ have been
performed by OPAL~\cite{OPAL}. 

%
\section{The ALEPH detector}
A detailed description of the ALEPH detector can be found in Ref.~\cite{Alnim},
and an account of its performance as well as a description of the
standard analysis algorithms can be found in Ref.~\cite{Alperf}.
Only a brief overview is given here.

Charged particles are detected in a magnetic spectrometer
consisting of a silicon vertex detector (VDET), a drift chamber (ITC)
and a time projection chamber (TPC), all immersed in a
1.5~T axial magnetic field provided by a superconducting solenoidal coil.
Between the TPC and the coil, a highly granular electromagnetic
calorimeter (ECAL) is used to identify electrons and photons and to
measure their energy. Surrounding the ECAL is the return yoke for the
magnet, which is instrumented with streamer tubes to form the hadron
calorimeter (HCAL). Two layers of external streamer tubes are used 
together with the HCAL to identify muons. 

The region near the beam line is covered by two luminosity calorimeters,
the SICAL and the LCAL. The SICAL provides coverage from 34 to 63~mrad from
the beamline while the LCAL provides coverage out to 160~mrad. The low angle
coverage is completed by the HCAL, which occupies a position behind the
LCAL and extends down to 106~mrad. The LCAL consists of two halves which 
fit together around the beamline; the area where the two halves come together 
is a region of reduced sensitivity. This ``vertical crack'' accounts for
only 0.05\% of the total solid angle coverage of the ALEPH detector.

The information obtained from the tracking system is combined 
with the information obtained from the calorimeters to form a 
list of ``energy flow particles''~\cite{Alperf}. These objects are used to
calculate the variables that are used in the analyses described
in Section~4.  
\section{Monte Carlo Simulation}
In the simulation of a stop signal, 
the most significant issues to be addressed
are the treatment of the stop
perturbative gluon radiation,
hadronisation and decay.

Since the stop is a scalar particle, the spectrum of gluon emission
differs from that of a quark. The standard shower evolution
programs would therefore need modifications to 
include the gluon emission from a spin-zero
particle. However, as pointed out in Ref.~\cite{LEP2phys},
the difference between the average energy loss due to perturbative
gluon emission off a spin-0 and a spin-1/2 particle
is small ($\lesssim$ $10^{-3}$) and can safely be
neglected within the approximations used by most shower Monte Carlo codes.

The stop lifetime is longer than the typical hadronisation time of
${\it{O}}(10^{-23} \ \rm{s})$, which corresponds to a width of
${\it{O}}(0.1 \ \rm{GeV})$. Stops therefore
hadronise into colourless ($\stop\bar{q}$) or
($\stop qq$) bound states before decaying.
This is incorporated in the generator by letting
stops hadronise as if they were ordinary quarks 
according to the LUND string fragmentation scheme
implemented in JETSET~7.4~\cite{JETSET}. 
A Peterson fragmentation function~\cite{Peterson} is used to describe the
stop fragmentation.
The $\epsilon_{\stop}$ parameter in the function is scaled
from b quarks following  the relation   
$\epsilon_{\stop} = \epsilon_{\rm{b}}m^2_{\rm{b}}/m^2_{\stop}$~\cite{Peterson},
with $\epsilon_{\rm{b}}$ = 0.0035~\cite{frag} and $m_{\rm{b}}$ = 5~$\gev$. 
Stop hadrons then decay according to a spectator model.
The effective spectator quark mass $\mx$, which takes into account 
non-perturbative effects, is set to 0.5~$\gev$. The
decay quark, c or b depending on the decay channel,
is allowed to develop a parton shower to take into account
hard gluon emission. At the end of the parton shower, a string 
is stretched among all coloured particles.  

A similar procedure is followed for the sbottom generator, taking into
account the fact that the $\sbot$ lifetime is much shorter than the 
$\stop$ lifetime. Depending on the $\sbot$ and $\neu$ mass difference and
coupling, the $\sbot$ can decay either before or after hadronisation.
Two sets of $\sbot$ signal samples, one
for each of these possibilities, were generated over the same range
of mass differences.


Signal samples 
were generated at $\rts$ = 130, 136, 161, and 172~GeV 
for various $(m_{\stop},m_{\neutr})$, $(m_{\sbot},m_{\neutr})$ or 
$(m_{\stop},m_{\sneut})$ masses. 
In these generations the mixing angle $\thetamix$ or $\thetab$ 
was set to zero; the selection efficiency depends on the value
of the mixing angle, since changing its value changes the
spectrum of initial state radiation.
Two sets of $\stop \to \rm{b} \ell \tilde{\nu}$ samples
have been produced. The first set assumes  equal branching
fractions for the $\stop$ decay to $\mathrm{e}$, $\mu$ or $\tau$, while
the second set assumes a branching fraction of 100\% for the decay to
$\tau$. All of these samples were processed though the full ALEPH
detector simulation.

The dependence of the selection efficiencies on the fragmentation
parameters and on the mixing angle is discussed in Section~5. 
The effect of the short $\sbot$ lifetime on the $\sbot$ selection efficiency
is also discussed in Section~5.

Monte Carlo samples corresponding to integrated luminosities at least 100
times that of the data have been fully simulated for 
the annihilation processes
$\rm{e^+ e^-} \to {\it f\bar{f}}$ and the 
various processes leading to four-fermion final states
($\mathrm{e}^+ \mathrm{e}^- \to \ww$,
$\mathrm{e}^+ \mathrm{e}^- \to \ewnu$,
$\mathrm{e}^+ \mathrm{e}^- \to \zee$ and
$\mathrm{e}^+ \mathrm{e}^- \to \zz$).
The two-photon processes $\gamma\gamma\to \rm{\ell^+ \ell^-}$
were simulated with an integrated luminosity about 20 times
that of the data, while the two-photon processes
$\gamma\gamma\to q \bar{q}$
were simulated with an integrated luminosity about three times
that of the data.
\section{Analysis}
Data collected at 
$\sqrt{s}$ = 130, 136, 161, 170, and 172~GeV have been analysed, 
corresponding to integrated luminosities of 
2.9, 2.9, 11.1, 1.1, and 9.5~$\rm{pb}^{-1}$, respectively.
To account for the dependence on
$\rts$, all cuts are performed in terms of variables normalised to the
beam energy.

\par

Two analyses are used to search for $\stop$ production.
The first one is sensitive to the decay 
\mbox{$\mathrm \stop \rightarrow c \neu$}
while the second one is sensitive to the decay 
$\mathrm \stop \rightarrow b \ell \snu$. 
Both channels are characterised
by missing momentum and energy.
The experimental topology depends largely on $\deltm$, the mass difference 
between the $\stop$ and the $\neu$ or $\snu$. When $\deltm$ is large,
there is a substantial amount of energy available for the visible system
and the signal events tend to look like $\ww$, 
$\ewnu$, $\zz$, and $\qqg$ events.
These processes are characterised by high multiplicity and high 
visible mass $M_{\mathrm{vis}}$.
When $\deltm$ is small, the energy available for the visible
system is small and the signal events are therefore similar to $\ggqq$ events.
The process $\ggqq$ is characterised by low multiplicity, 
low $M_{\mathrm{vis}}$, low total transverse momentum $\pt$ 
and the presence of energy near the beam axis. In order to cope with
the different signal topologies and background situations, 
each analysis employs a low $\deltm$ selection and a high $\deltm$
selection.

\par

The values of the analysis cuts are set in an unbiased way following the 
$\bar{N}_{95}$ procedure~\cite{N95}. In this procedure, 
the cut values are varied and applied to the
background samples and the signal samples in order to calculate
$\mathrm{\bar{\sigma}_{95}}$, the expected 95\% Confidence Level (C.L.) 
limit on the signal cross section. The final cut values used in the
analyses are the ones which minimise $\mathrm{\bar{\sigma}_{95}}$.
Cuts used to eliminate background from $\ggqq$ events are
not varied. Such events are difficult to simulate when they go 
into the low angle region of the detector. 
Conservatively, the values of the cuts used against $\ggqq$ 
events are tighter than the values given by the 
$\bar{N}_{95}$ procedure.


\par

The experimental topology of the process 
$\mathrm{e^{+}e^{-}\rightarrow \sbot \bar{\sbot}}$ 
($\sbot \rightarrow \mathrm{b}\neu$)
is quite similar to that of the process 
$\mathrm{e^{+}e^{-}\rightarrow \stop \bar{\stop}}$ 
($\stop \rightarrow \mathrm{c}\neu$).
A common selection is therefore used to search for these two processes.
\subsection{Search for $\mathrm \stop \rightarrow c\neu$ and
$\mathrm \sbot \rightarrow b\neu$}

\begin{figure}[tcb]
\begin{center}
\begin{tabular}{cc}
\epsfig{file=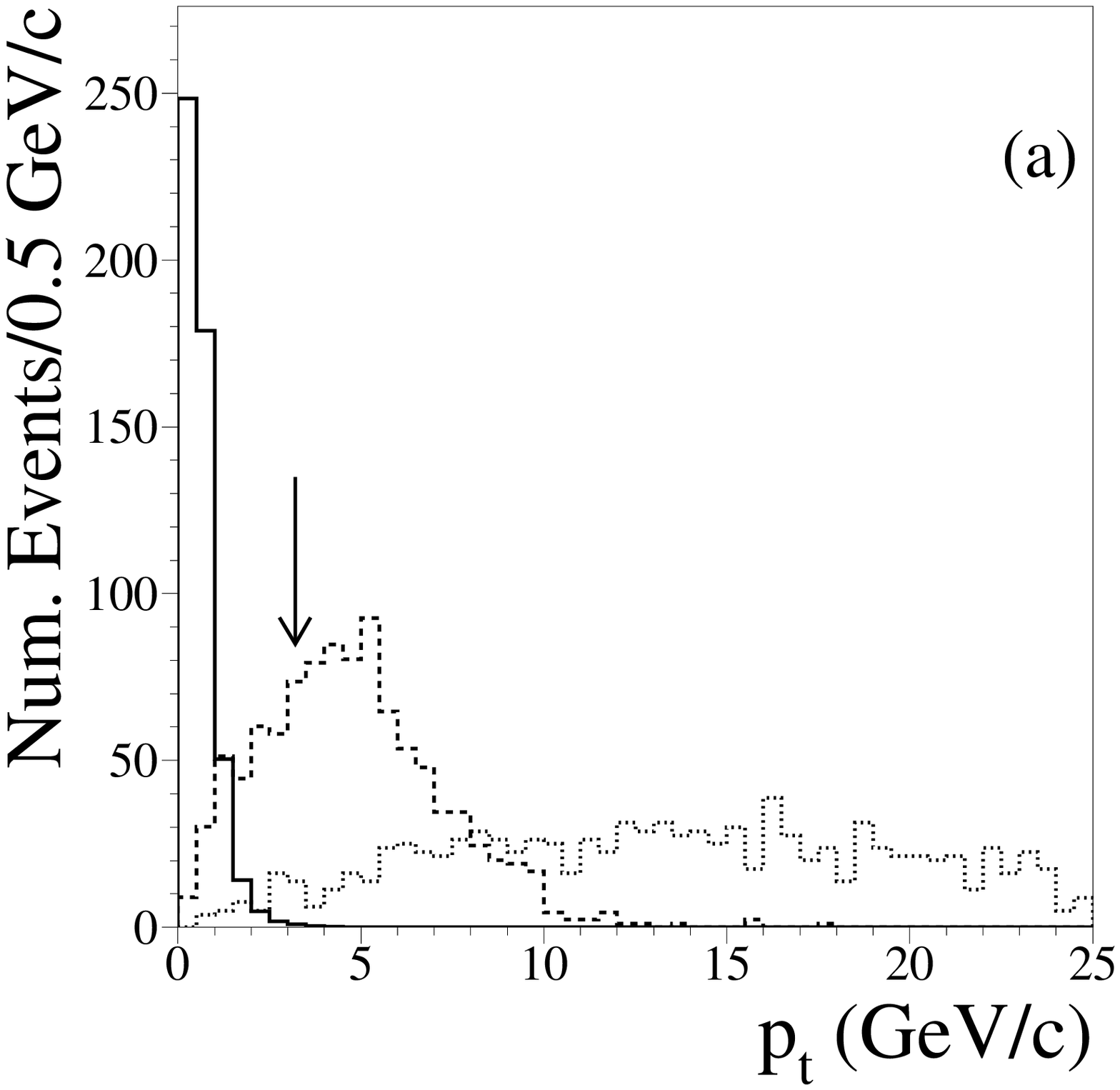,width=0.46\textwidth}
\epsfig{file=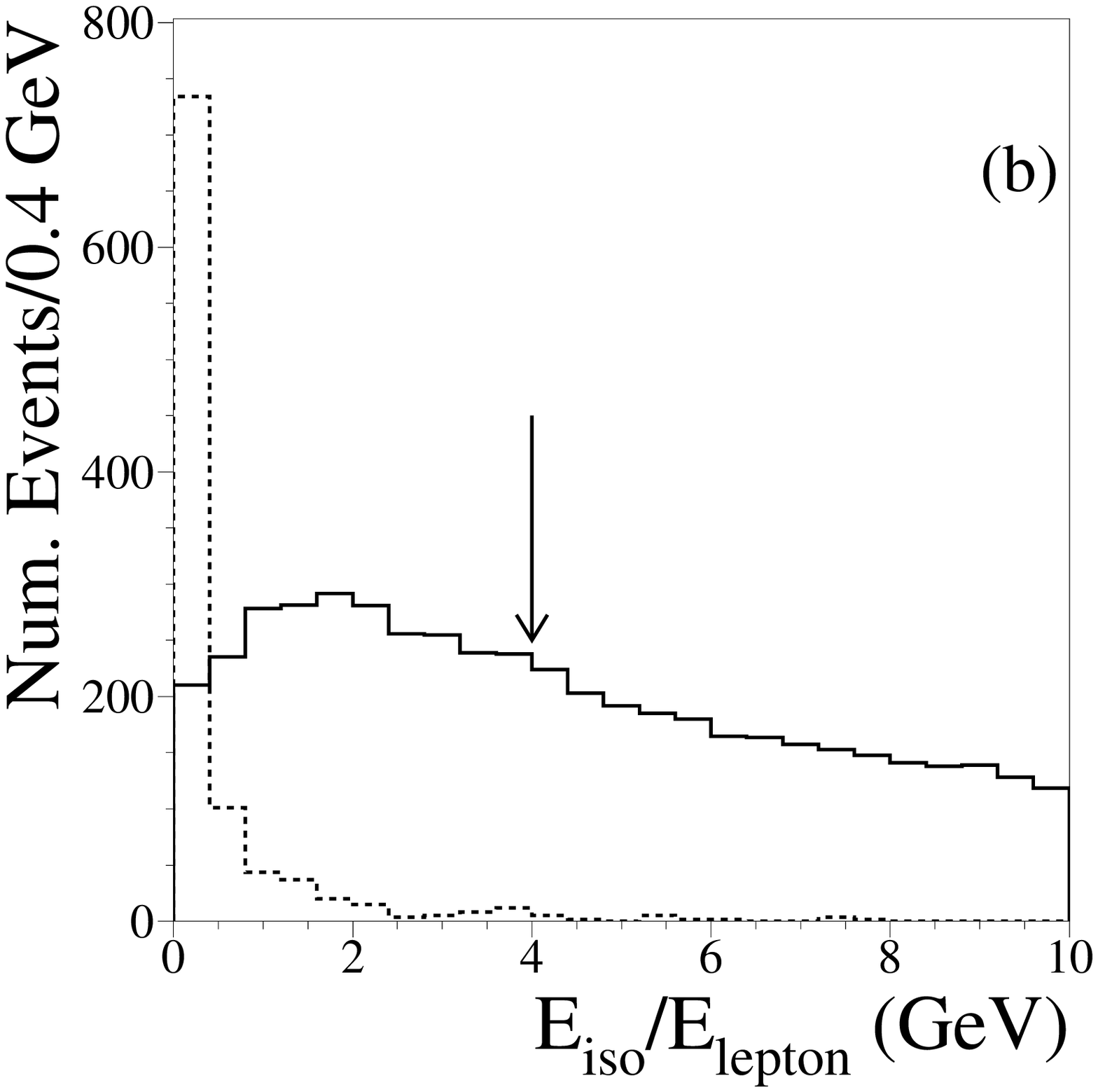,width=0.46\textwidth}
\end{tabular}
\end{center}
\caption{\rm
(a) $\pt$ for $\ggqq$ and
    $\stop \rightarrow \mathrm{c}\neu$ at $\rts$ = 161~GeV. 
    The solid histogram gives the
    $\ggqq$ distribution, the dashed histogram gives the signal distribution
    for \mbox{$\mstop$ = 65 $\gev$} and
    $\deltm$ = 5 $\gev$, the dotted histogram gives the signal distribution
    for \mbox{$\mstop$ = 65 $\gev$} and $\deltm$ = 15 $\gev$.
    The cut $\pt$ $>$ 2\%$\rts$ is indicated by the arrow.
(b) $E_{\mathrm{iso}}$/$E_{\mathrm{lepton}}$ for $\qqg$ and 
    $\stop \to \rm{b} \ell \tilde{\nu}$ at $\rts$ = 161~GeV. 
    The solid histogram gives the $\qqg$
    distribution, the dashed histogram gives the signal distribution for 
    \mbox{$\mstop$ = 60 $\gev$} and $\deltm$ = 20 $\gev$.
    The cut $E_{\mathrm{iso}}$/$E_{\mathrm{lepton}}$ $<$ 4 is indicated by the
    arrow. In (a), the cut $\elow$ = 0 has been applied. In (b), at least
    one identified electron or muon is required. Normalization 
    for the plots is arbitrary.
\label{stdist}}
\end{figure}

The processes $\mathrm e^{+}e^{-} \rightarrow \stop\bar{\stop}$ 
($\mathrm \stop \rightarrow c\neu$) and
$\mathrm e^{+}e^{-} \rightarrow \sbot\bar{\sbot}$
($\mathrm \sbot \rightarrow b\neu$)
are characterised by two acoplanar jets and missing mass and energy. 
Two selections are employed, one for the small $\deltm$
case \mbox{($\deltm$ $<$ 10~$\gev$)} and one for the large $\deltm$ case
\mbox{($\deltm$ $\geq$ 10~$\gev$)}. 
A common preselection is used against $\ggqq$ 
events in both the low and high $\deltm$ analyses.
The number of charged particle tracks $\nch$ must be at least four, 
$M_{\mathrm{vis}}$ must
be larger than 4~$\gev$ and $\pt$ (Figure~\ref{stdist}a) 
must be larger than 2\%$\rts$, or 4\%$\rts$ 
if the missing momentum points to within $\mathrm 15^{\circ}$ in azimuth
from the vertical crack in LCAL. 
The polar angle of the 
missing momentum vector, $\thmiss$, must 
be greater than $18^{\circ}$ and the energy detected
within $12^{\circ}$ of the beam axis, $\elow$, 
must be less than 5\%$\rts$. Both the acoplanarity and the 
transverse acoplanarity must be less than $175^{\circ}$.
The acoplanarity is defined to be $180^{\circ}$ for a back-to-back
topology and is calculated from the momenta directions of the two event
hemispheres, defined by a plane perpendicular to the thrust axis.
The transverse acoplanarity is obtained by projecting the event onto a
plane perpendicular to the beam axis, then calculating the two-dimensional
thrust axis and dividing the event into two hemispheres by a plane 
perpendicular to that thrust axis.
Both of these cuts are also effective against $\qqg$ background.

\subsubsection{Low $\deltm$ selection}
Most of the cuts in the low $\deltm$ analysis are designed to
eliminate the remaining background from $\ggqq$ events.  
The $\pt$ cut is reinforced by calculating $\pt$ 
excluding the neutral hadrons found by the energy flow algorithm 
and requiring it to be
greater than 2\%$\rts$. The $\pt$ is also calculated  
with only the charged particle tracks and 
required to be greater then 1\%$\rts$.
These cuts eliminate $\gg$ events that have a large $\pt$ due to 
spurious calorimeter objects; these objects can occur when soft
tracks are not correctly associated with deposits in the ECAL or HCAL.  
Such events are also eliminated by asking that the most energetic neutral 
hadronic deposit be less than 30\% of the total visible energy 
$E_{\mathrm{vis}}$. 
To eliminate $\gg$ events that pass the $\pt$
cuts, $\elow$ must be equal to zero, 
$\thmiss$ must be greater than $37^{\circ}$, $\theta_{\mathrm{thrust}}$,
the polar angle of the thrust axis, must be greater than $41^{\circ}$, and
the missing mass $M_{\mathrm{miss}}$ divided by $M_{\mathrm{vis}}$ must be
less than 25.
Also of use is the fact that the missing momentum in $\ggqq$ and $\qqg$ events 
can arise from neutrinos produced in semileptonic decays.
When these decays occur within a jet, the resulting missing $\pt$ 
is not isolated. Signal events are therefore selected by requiring the    
energy $E_{\mathrm{w}}$ in a $30^{\circ}$ azimuthal wedge 
around the direction of missing $\pt$ to be less than 25\%$\rts$.

\par

If a scattered electron from a $\ggqq$
process goes into an insensitive region of the detector, only a
small fraction of the electron energy may be recorded. The missing 
electron energy can lead to a large measured $\pt$, faking a signal.  
These fake signals can be eliminated by calculating the
scattered electron angle $\thscat$ from the 
$\pt$, assuming the other electron to be undeflected, and by computing
the angle $\diffa$ between the calculated electron direction 
and the closest energy deposit. 
The fake signals surviving the $\pt$ cut 
usually have a large value of $\thscat$, because the
$\pt$ imbalance is large, and a small 
value of $\diffa$, because the calculated
electron direction points to the energy deposit from the scattered
electron. Both $\thscat$ and $\diffa$ are incorporated into 
the analysis through the cut $\diffa > 60^{\circ} - 10 \times \thscat$.

\par

Additional cuts are used against the $\ggtt$ background. Most of the $\ggtt$
events that survive the above cuts have four charged particle tracks from
the decays $\tau\rightarrow \mbox{one-prong}$, 
$\tau\rightarrow \mbox{three-prong}$, and 
the low visible mass and high value of 
acoplanarity characteristic of $\gg$ events
in general. In order to eliminate these events, any four-track event 
must have transverse acoplanarity less than $150^{\circ}$
or visible mass greater than 20~$\gev$. As an additional safeguard,
all four-track events are required to have a visible mass larger than 
8~$\gev$ regardless of the value of the transverse acoplanarity.

\par

The low $\deltm$ analysis is completed by applying cuts against low mass
$\ww$, $\zz$, and $\ewnu$ events. A cut of thrust $<$ 0.97 is effective 
against $\zz$ (with $\mathrm{Z} \rightarrow \nu\bar{\nu}$), 
while $\ww$ and $\ewnu$ events are eliminated by requiring
that $E_{\mathrm{vis}}$ be less than 26\%$\rts$. 
Events from the process $\mathrm
\ww \rightarrow \ell\nu_{\ell} \tau\nu_{\tau}$, where 
the $\tau$ subsequently undergoes a three-prong decay, are eliminated 
by requiring that the event mass excluding identified 
electrons and muons be greater than 3~$\gev$.

\subsubsection{High $\deltm$ selection}

The main background in the high $\deltm$ case comes from $\ww$,
$\ewnu$, $\zz$, and $\qqg$. Events from $\gg$ processes may still contribute
to the background because they have a very large cross section
and because detector effects may lead to extreme values for 
variables such as $\pt$. 
Background from $\gg$ is reduced by requiring that $\nch$ be  
greater than six 
and that $\pt$ be greater 
than 5\%$\rts$, or 7.5\%$\rts$ if the missing momentum points 
to within $15^{\circ}$ of the vertical LCAL crack. Additional $\gg$ 
events are removed by requiring that $\pt$ be 
greater than 20\%$E_{\mathrm{vis}}$. As in 
the low $\deltm$ selection, it is necessary to guard against $\gg$ events 
that have a large $\pt$ due to a missed association between soft tracks 
and calorimetry deposits. This is done by demanding that the total energy
from neutral hadrons be less than 30\%$E_{\mathrm{vis}}$; 
this is relaxed to 45\%$E_{\mathrm{vis}}$
if the $\pt$ calculated without neutral hadrons 
is greater than 3\%$\rts$. 
Other cuts which are effective against  $\gg$ events are $\diffa$ $>$ 
$5^{\circ}$, $E_{\mathrm{w}}$ $<$ 7.5\%$\rts$ 
and the total energy more 
than $30^{\circ}$ away from the beam greater than 30\%$E_{\mathrm{vis}}$.

\par

Finally, cuts against $\ww$, $\ewnu$, and $\zz$ are applied. Events
from $\zz$ are eliminated by requiring that the thrust be less than 0.935.
To eliminate $\ww$ events in which one of the W's decays leptonically,
any identified electron or muon must have an energy less than 20\%$\rts$. 
In order to further reduce background from $\ww$ and $\ewnu$ events,
an upper cut is applied on the visible mass.
The optimal value of this cut as determined by the 
$\bar{N}_{95}$ procedure depends on the mass difference of the signal 
sample considered. 
A hypothesis of $\deltm$ = 15~$\gev$ gives an optimal
value of 0.315$\rts$ for the $M_{\mathrm{vis}}$ cut while a hypothesis
of $\deltm$ $\geq$ 35~$\gev$ gives an optimal value of 0.375$\rts$ for the 
$M_{\mathrm{vis}}$ cut.
 

\par

The high $\deltm$ selection changes as a function of $\deltm$ through the
$M_{\mathrm{vis}}$ cut. When this selection is applied to the data, the
loosest $M_{\mathrm{vis}}$ cut is used. In the case that limits must
be set, a candidate is counted for a given value of $\deltm$ only if
it has a visible mass less than the $M_{\mathrm{vis}}$ cut used for that
value of $\deltm$.

\subsubsection{Selection efficiency and background}

\begin{figure}[tcb]
\begin{center}
\begin{tabular}{cc}
\epsfig{file=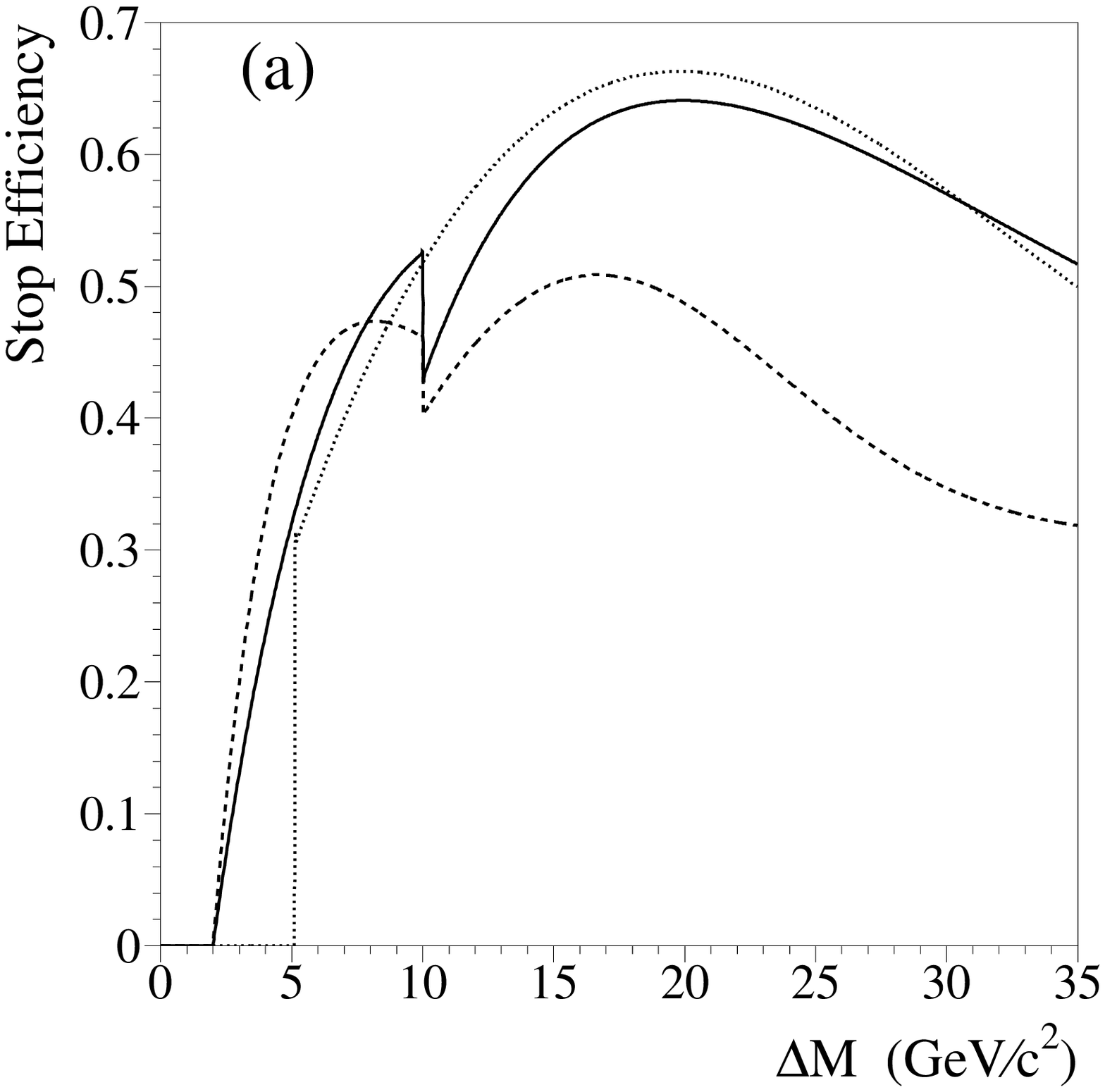,width=0.46\textwidth} &
\epsfig{file=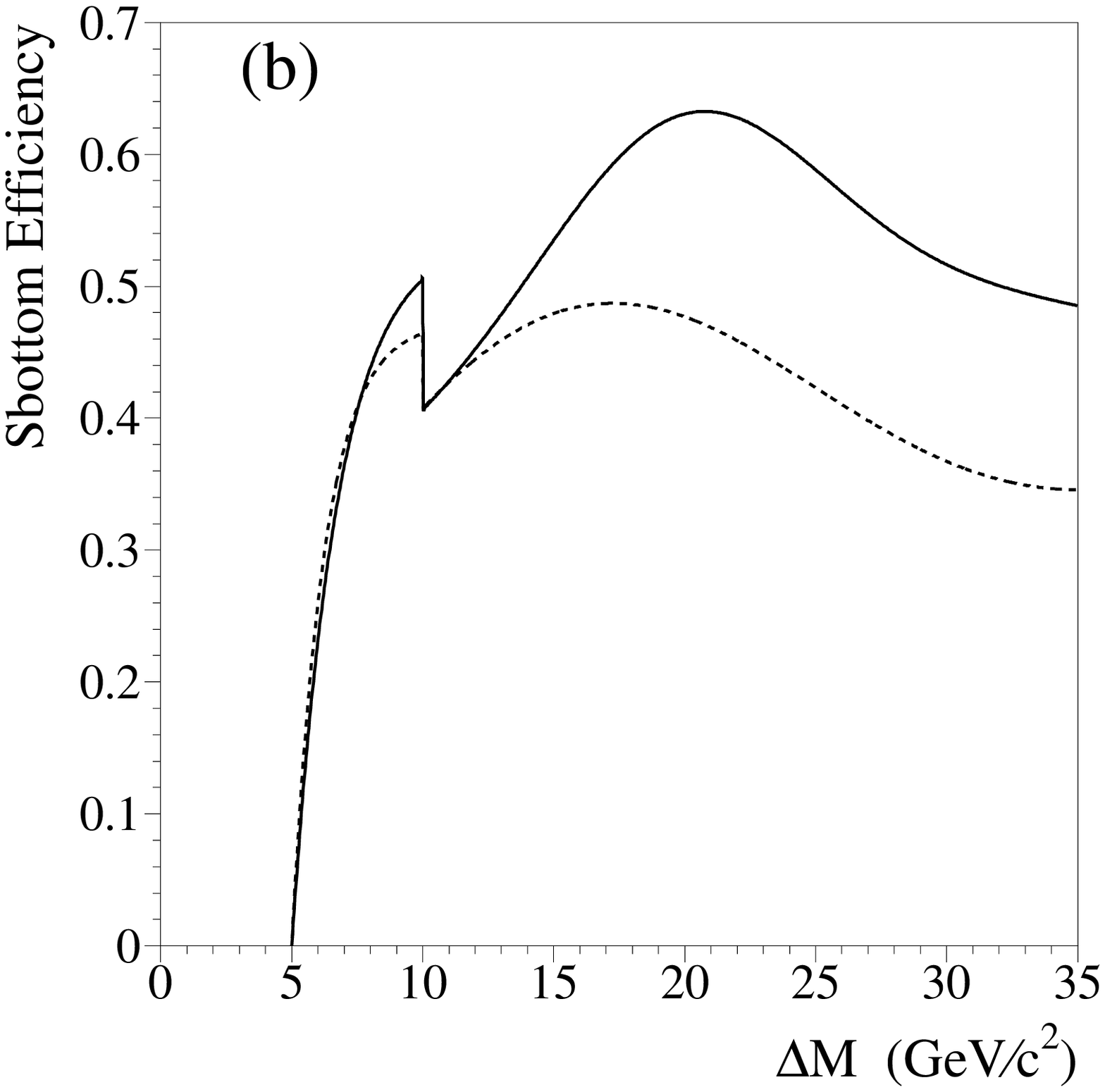,width=0.46\textwidth}
\end{tabular}
\end{center}
\caption{\rm Efficiencies as a function of $\deltm$. (a)
Efficiency for a 65 $\gev$ stop decaying as
$\stop \rightarrow \mathrm{c}\neu$ (solid curve), a
50 $\gev$ stop decaying as $\stop \rightarrow 
\mathrm{c}\neu$ (dashed curve)
and a 60~$\gev$ stop decaying as 
$\stop \rightarrow \mathrm{b}\ell\tilde{\nu}$
(dotted curve). (b) Efficiency for a 60 $\gev$ sbottom
(solid curve) and a 50 $\gev$ sbottom (dashed curve) decaying as
$\sbot \rightarrow \mathrm{b}\neu$.
\label{steff}}
\end{figure}

To combine the 
low and high $\deltm$ selections, three possibilities are considered:
the low $\deltm$ selection may be used, the high $\deltm$ selection
may be used, or both selections may be used.
According to the $\bar{N}_{95}$ procedure   
the two selections should not
be used simultaneously for any value of $\deltm$. 
For $\deltm$ $<$ 10~$\gev$, the low $\deltm$ selection
is used, while for $\deltm$ $\geq$ 10~$\gev$, the high $\deltm$
selection is used. The $\stop$ efficiencies 
are shown in Figure~\ref{steff}a while the $\sbot$ efficiencies are
shown in Figure~\ref{steff}b. These $\sbot$ efficiencies are evaluated 
assuming that the $\sbot$ hadronises before it decays. 

For the low $\deltm$ selection,
the requirement that $\elow$ = 0 results in an inefficiency
due to beam-related and detector 
background. The size of this effect ($\sim$ 4\%) has been measured 
using events triggered at random beam crossings and the
low $\deltm$ selection efficiency is decreased accordingly. 

The background to the low $\deltm$ selection is dominated by
$\ggqq$ and $\ggtt$ and has a total expectation of
0.9~events \mbox{($\sim$ 40 fb)} at $\rts$ = 161--172~GeV 
and 0.2~events \mbox{($\sim$ 30 fb)} at $\rts$ = 130--136~GeV.
For the high $\deltm$ selection, the background is dominated by
$\ww$, $\ewnu$, $\zz$, and $\qqg$ at $\rts$ = 161--172~GeV and by $\qqg$ at
\mbox{$\rts$ = 130--136~GeV.} The total 
background expectation for this selection
is 1.0 event \mbox{($\sim$ 50 fb)} at \mbox{$\rts$ = 161--172~GeV} 
and 0.2 events \mbox{($\sim$ 30 fb)} at 
$\rts$ = 130--136~GeV, using the loosest value of the 
$M_{\mathrm{vis}}$ cut.


%

%
%
%
%
\subsection{Search for  $\stop \to \rm{b} \ell \snu$}
 The experimental signature for $\stop \to \rm{b} \ell \snu$ is 
 two acoplanar jets plus two leptons with missing momentum.
 The leptons tend to have low momenta, especially for 
 low $\deltm$ signals; when
 $\deltm$ is 8~$\gev$, the most energetic lepton often has a 
 momentum between 1 and 2~$\gvm$. 
 In order to identify electrons and muons, loose identification 
 criteria based on the pattern of deposits in the ECAL and the HCAL
 have been applied. These loose criteria allow 1~$\gvm$ electrons and
 1.5~$\gvm$ muons to be identified. Since low-momenta lepton candidates
 are often mis-identified pions, other analysis cuts must be 
 used to keep the background at a low level. 
 Two selections are used, one for the small $\deltm $
 case ($\deltm$ $<$ 10~$\gev$) and the other for the large $\deltm $ case 
 ($\deltm$ $\geq$ 10~$\gev$). A preselection common to both the low and high
 $\deltm$ selections is used against the $\ggqq$ background. 
 It is required that $\nch$ be greater than six  
 and $M_{\mathrm{vis}}$ be greater than 
 8\%$\rts$. It is also required  
 that $\pt$ be greater than 1.25\%$\rts$, 
 $\elow$ be smaller than 2~GeV, and
 $\diffa$ be greater than $50^{\circ} - 20 \times \thscat$.
 In order to eliminate the radiative $f\bar{f}\gamma$
 events in which a return to the Z peak has occurred, events with
 a longitudinal momentum greater than 30\%$\rts$ are rejected.

\subsubsection{Low $\deltm$ selection}

 If $\deltm$ is small the visible energy is also small
 and both the jets and leptons are very soft. Since very soft
 leptons might not be identified, events with no electrons or 
 muons are accepted.
 The main background arises from $\ggqq$. It is therefore required that 
 \mbox{$\elow$ = 0} and that both $\thmiss$ and  
 $\theta_{\mathrm{thrust}}$
 be greater than  $37^{\circ}$. An acoplanarity between $100^{\circ}$ 
 and $179^{\circ}$ is also required. 
 There must be at least one electron or muon with momentum greater 
 than 1\%$\rts$, otherwise  
 both the $\pt$ cut and the two-dimensional cut in the  $\diffa$-$\thscat$ 
 plane are tightened: $\pt$ $>$ 2\%$\rts$, 
 $\diffa > 115^{\circ} - 20 \times \thscat$.
\par
 The $\ww$ background is eliminated by requiring that the missing mass be 
 greater than 82.5\%$\rts$ and that the hadronic mass be smaller than 
 5\%$\rts$ if at least one electron or muon is identified.
 The $\qqg$ events are rejected by requiring that the thrust be
 smaller than 0.9.

 \par

\subsubsection{High $\deltm$ selection}

 For large mass differences at least  one 
 electron or muon with  momentum between 2 and 35~$\gvm$ is required.
 It is further required that $E_{\mathrm{iso}}$,
 the energy in a $30^{\circ}$ cone around the direction of the electron
 or muon momentum (Figure~\ref{stdist}b), be smaller than 
 four times the electron or muon energy.
 If a second electron or muon is identified, $E_{\mathrm{iso}}$ 
 is required to be smaller than 10 times the electron or muon energy.
 If only one electron or muon is found, 
 a tau jet is selected using the JADE algorithm with 
 $y_{\mathrm{cut}}$ = 0.001.
 This candidate $\tau$ jet must have
 an energy smaller than 30~GeV, have less than 
 2~GeV of energy carried by neutral hadrons, and have an angle of at least
 $20^{\circ}$ with the nearest jet. Finally,
 the missing mass is required to be greater than 25\%$\rts$.
\par
 To reinforce the $\ggqq$ rejection further cuts are needed.  
 It is required that  $\thmiss$  be greater than 
 $18^{\circ}$, that the transverse acoplanarity be smaller than 
 $176^{\circ}$ and that the acollinearity be smaller than $174^{\circ}$.
 If only one electron or muon is identified the 
 hadronic neutral mass must be smaller than 30\%$E_{\mathrm{vis}}$ 
 and the cuts on  $\thmiss$ and $\pt$ are tightened:  
 $\thmiss$ $>$ $26^{\circ}$, $\pt$ $>$ 3\%$\rts$.
\par
 The $\ww$ background events are 
 eliminated by requiring
 that $M_{\mathrm{vis}}$  be
 smaller than 
 74\%$\rts$ and that the hadronic 
 mass be less than 37\%$\rts$. 
 It is also required that the quadratic mean  of the two 
 inverse hemisphere boosts 
 ($\sqrt{((m_1/E_1)^2+(m_2/E_2)^2)/2}$ with $m_{1,2}$
 and $E_{1,2}$ the two hemisphere masses and energies) be greater than 
 0.2. The remaining $\qqg$ background is reduced by requiring 
 that the thrust be smaller than 0.925.

\subsubsection{Selection efficiency and background}

 The low and high $\deltm$ selections are combined using the same
 procedure as in Section~4.1.3. In contrast to the situation  
 for the $\stop \rightarrow \mathrm{c}\neu$ channel, the smallest value of
 $\mathrm{\bar{\sigma}_{95}}$ is obtained when
 the low and high $\deltm$ selections are used
 simultaneously. This is true for all values of $\deltm$. 
 Shown in Figure~\ref{steff}a is the efficiency assuming equal
 branching fractions for the $\stop$
 decay to $\mathrm{e}$, $\mu$ or $\tau$. If the branching ratio
 to $\tau$ is 100\%, the efficiency is about 35\% for a $\deltm$
 between 10 and 35 $\gev$.
 As is the case for the
 $\stop \rightarrow \mathrm{c}\neu$ channel, the inefficiency caused
 by the beam-related and detector background is taken into account.

Most of the background comes from the high $\deltm$ selection and is
dominated by $\qqg$ at 
\mbox{$\rts$ = 130--161~GeV} and by $\ww$ and $\qqg$ at
$\rts$ = 170--172~GeV. A total of 0.8~events 
\mbox{($\sim$ 30 fb} at 161~GeV and $\sim$ 50 fb at 172~GeV) are expected at 
$\rts$ = 161--172~GeV while 0.2~events 
\mbox{($\sim$ 30 fb)} are expected at $\rts$ = 130--136 GeV.


%
%
%
\section{Systematic Uncertainties}

The systematic uncertainty on the $\stop$ and $\sbot$ selection 
efficiencies comes mainly  from the limited knowledge 
of $\stop$ and $\sbot$ physics (hadronisation and decay). Uncertainties
related to detector effects, to the size of the signal samples, and to 
the parameterisation of the signal efficiencies are 
also considered, and for the $\stop \to \rm{b} \ell \snu$  
analysis the effects of lepton identification are taken 
into account. The physics model used in the generators is described in 
Section 3; the systematic effects are studied by varying the parameters 
of the model and checking the resultant effect on the efficiency.


\par

The change in the efficiency due to the systematic effects is 
shown in Table~3. When $\deltm$ is small, the uncertainties associated 
with the $\stop$ and $\sbot$ physics are relevant. The largest change in the 
low $\deltm$ efficiency comes from the variation in $\mx$.
This variation changes the invariant mass
available for the hadronic system and thus the multiplicity and 
event shape. To quantify these effects, 
$\mx$ is varied from 0.3~$\gev$ to 1.0~$\gev$, a range 
much larger than that implied by low energy 
measurements. When $\deltm$ is large, the selection
efficiencies are insensitive to the values of the parameters, changing
by only $\sim$ 2\% relative even for $\mx$ = 2~$\gev$.  


The fragmentation parameters are varied over a range suggested by
LEP1 measurements. In the case of 
$\epsilon_{\stop}$ the error
is propagated from $\epsilon_{\mathrm{b}}$
according to the formula described in 
Section~3, and for the  $\stop \to \rm{b} \ell \snu$ channel
$\epsilon_{\mathrm{b}}$ is varied simultaneously with  $\epsilon_{\stop}$.
Similarly, for the $\sbot \to \rm{b} \neu$ channel 
$\epsilon_{\mathrm{b}}$ is varied simultaneously with  $\epsilon_{\sbot}$.  
For the large $\deltm$ case the fragmentation parameters are
varied more drastically, but even drastic variations have little effect
on the efficiency; for instance, 
when $\epsilon_{\stop}$ = $\epsilon_{\mathrm{b}}$,
the relative change in large $\deltm$ $\stop$ efficiencies is only $\sim$ 2\%.

The systematic effect of varying the mixing angles 
is quantified by evaluating
the efficiencies on a set of $\stop$ samples 
generated with $\thetamix$ = 56$^{\circ}$ and on a set of $\sbot$ samples 
generated with $\thetab$ = 68$^{\circ}$. 
For these values of mixing, the stops and sbottoms decouple from the 
$\mathrm{Z}$ and the change in efficiencies due to differing amounts
of initial state radiation is maximal.

\par

The structure of the matrix element~\cite{Hikasa} in the semileptonic 
decay $\stop \to \rm{b} \ell \snu$ is also considered. 
Two sets of $\stop \to \rm{b} \ell \snu$
signal samples are generated. One set includes the 
the matrix element, treated as in Reference~2, 
while the other set employs a phase
space decay model. Including the matrix element increases the 
efficiency of the $\stop \to \rm{b} \ell \snu$ selection
by about 5\% relative with respect to the 
phase space decay model. Conservatively, the
phase space decay model is used to obtain the 
$\stop \to \rm{b} \ell \snu$ efficiencies.

\par

The effect of the relatively short $\sbot$ lifetime has been checked
by comparing the two sets of $\sbot$ signal samples. 
Higher efficiencies are always obtained from the set in which the
$\sbot$ decays before hadronisation. 
The lower efficiencies, obtained under the assumption that the $\sbot$
hadronises before decay, are taken as the actual efficiencies; this 
helps ensure that any limits set on $\sbot$ production will be
conservative.
  
\par

\begin{table}
\caption{ 
  Summary of relative systematic uncertainties on the $\stop$ and $\sbot$ 
  selection efficiencies. The ranges of variation are those used for the
  low $\deltm$ case. 
 \label{stsys}}
 \begin{center}
 \begin{tabular}{|l|c|c|c|c|c|c|}           \hline
   \multicolumn{7}{|c|}{Systematic Uncertainties (\%)} \\ \hline
   \multicolumn{1}{|c|}{\em Type} &
   \multicolumn{2}{c|}{$\stop\to \rm{c}\neu$} &
   \multicolumn{2}{c|}{$\sbot\to \rm{b}\neu$} &
   \multicolumn{2}{c|}{$\stop \to \rm{b}\ell\snu$} \\ 
                                                      \hline\hline
 
         & High $\deltm$ & Low $\deltm$ & High & Low & High  & Low  \\
                                               \cline{2-7}

  $\mx$       (0.3--1.0 $\gev$)    & 3 & 10 &  4 & 11 & 3   &  15  \\

  $\epsilon_{\stop}, \epsilon_{\mathrm{b}}(\epsilon_{\mathrm{b}} 0.002-0.006)$ &  2 & 2 &  - & - &  2  & 2   \\

  $\epsilon_{\sbot}, \epsilon_{\mathrm{b}}(\epsilon_{\mathrm{b}} 0.002-0.006)$ &  - & - &  1 & 2   &  -  & -   \\

  $\epsilon_{\mathrm{c}}$ (0.02--0.06) & 3 & 7 &  -  & - & -   &  -  \\

  $\thetamix$ (0$^{\circ}$--56$^{\circ}$) 
   &   1 & 3      &     - & -      & 2   &  1  \\

  $\thetab$ (0$^{\circ}$--68$^{\circ}$) 
   &   - & -      &     3 & 2      & -   &  -  \\

  Monte Carlo statistics & 3 & 3 & 3 & 3 & 3 & 3 \\

   detector effects  & {\it negl.} & {\it negl.} & {\it negl.} & {\it negl.}  & 3   & 3    \\ \cline{1-7}

   TOTAL      &  6 & 13   & 6 & 12   & 6  & 16 \\ \hline

\end{tabular}
\end{center}
\end{table}


The size of the signal samples, 1000 events, leads to a 
relative uncertainty of less than
2\%, while the parameterisation of the signal efficiencies leads
to an additional relative uncertainty of $\sim$ 2\%. The total
statistical uncertainty associated with the Monte Carlo signal
simulation is therefore $\sim$ 3\% relative.

Detector effects have been studied for the variables used in the 
analyses. Events in the data from $\qqg$ final states are 
selected with a loose set of cuts 
and compared with the $\qqg$ Monte Carlo. 
All of the relevant variables, such as $\pt$ and $\diffa$, show
good agreement. The lepton
isolation and the lepton identification, 
which are crucial for the $\stop \to \rm{b} \ell \snu$ analysis,
are also considered. The lepton isolation
shows good agreement between $\qqg$ Monte Carlo and data,
while the lepton identification is found to lead to a
3\% systematic error.


\par

The systematic errors are incorporated into the final result using
the method described in Reference \cite{syse}.

\par

%
%
\section{Results}

%

\begin{figure}[p]
\begin{center}
\begin{tabular}{cc}
\epsfig{file=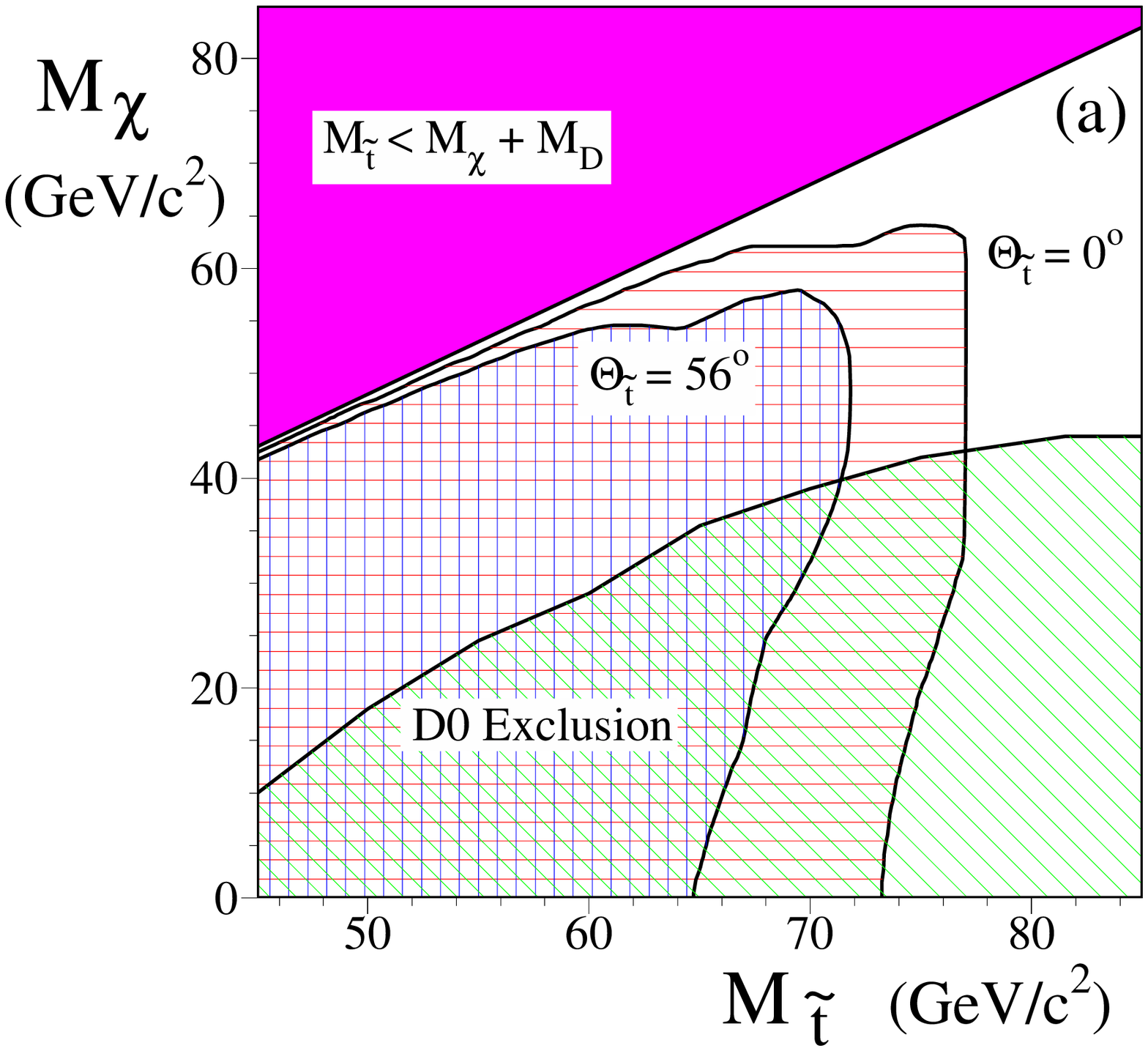,width=0.46\textwidth} &
\epsfig{file=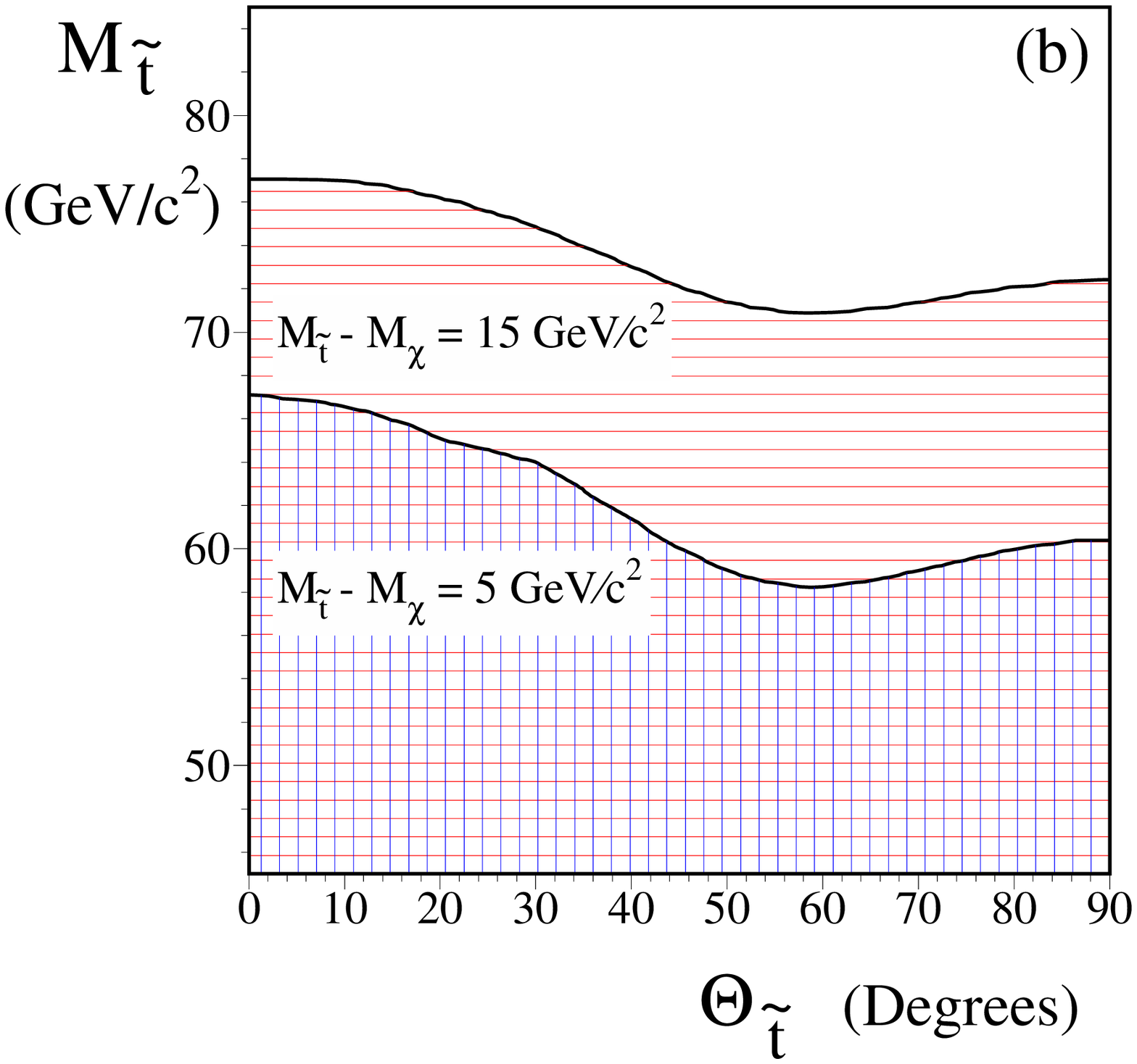,width=0.46\textwidth}
\end{tabular}
\epsfig{file=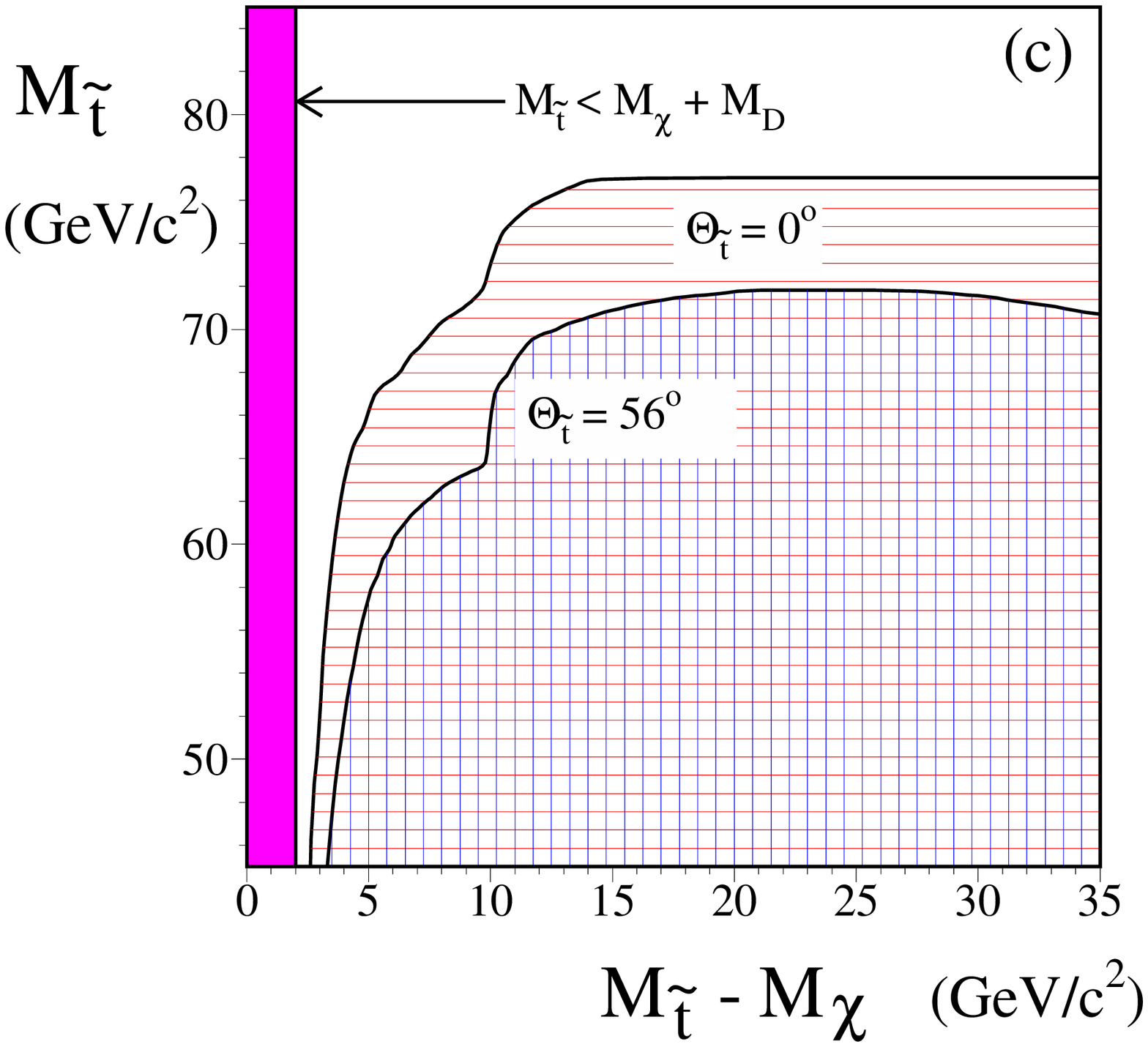,width=0.46\textwidth}
\end{center}
\caption{\rm Excluded regions assuming $\stop \rightarrow \mathrm{c}\neu$. 
(a) Excluded region in the $m_{\neu}$ vs 
$m_{\stop}$ plane, including the region excluded by the
D0 collaboration. (b) Excluded region in the  
$m_{\stop}$ vs $\mathrm \thetamix$ plane. (c) Excluded region in the  
$m_{\stop}$ vs $\deltm$ plane. In (a) and (c), excluded regions are given for
$0^{\circ}$, corresponding to the maximum $\stop$-Z coupling, and for
$56^{\circ}$, corresponding to the minimum $\stop$-Z coupling.
\label{stchi}}
\end{figure}
\begin{figure}[p]
\begin{center}
\begin{tabular}{cc}
\epsfig{file=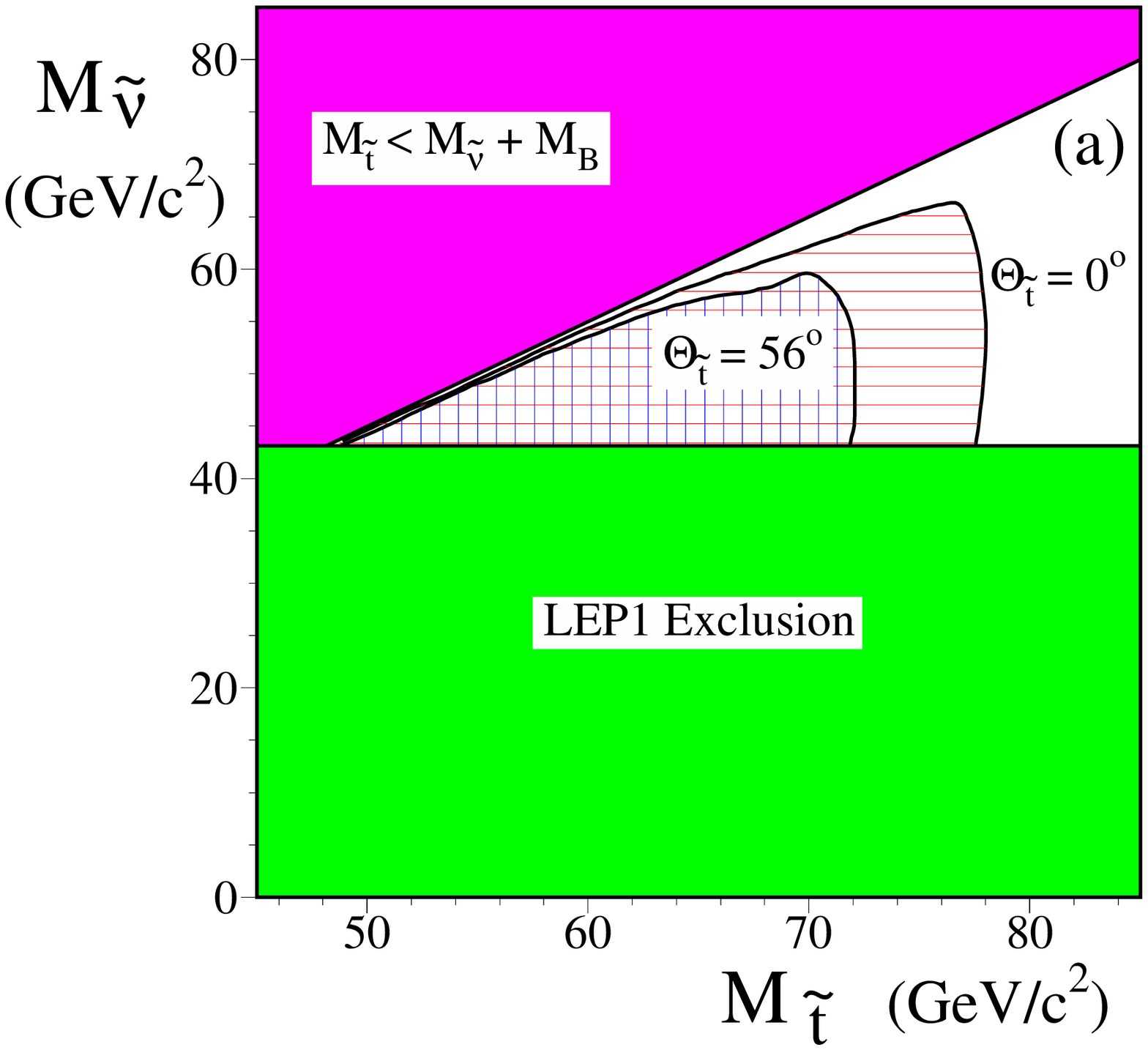,width=0.46\textwidth} &
\epsfig{file=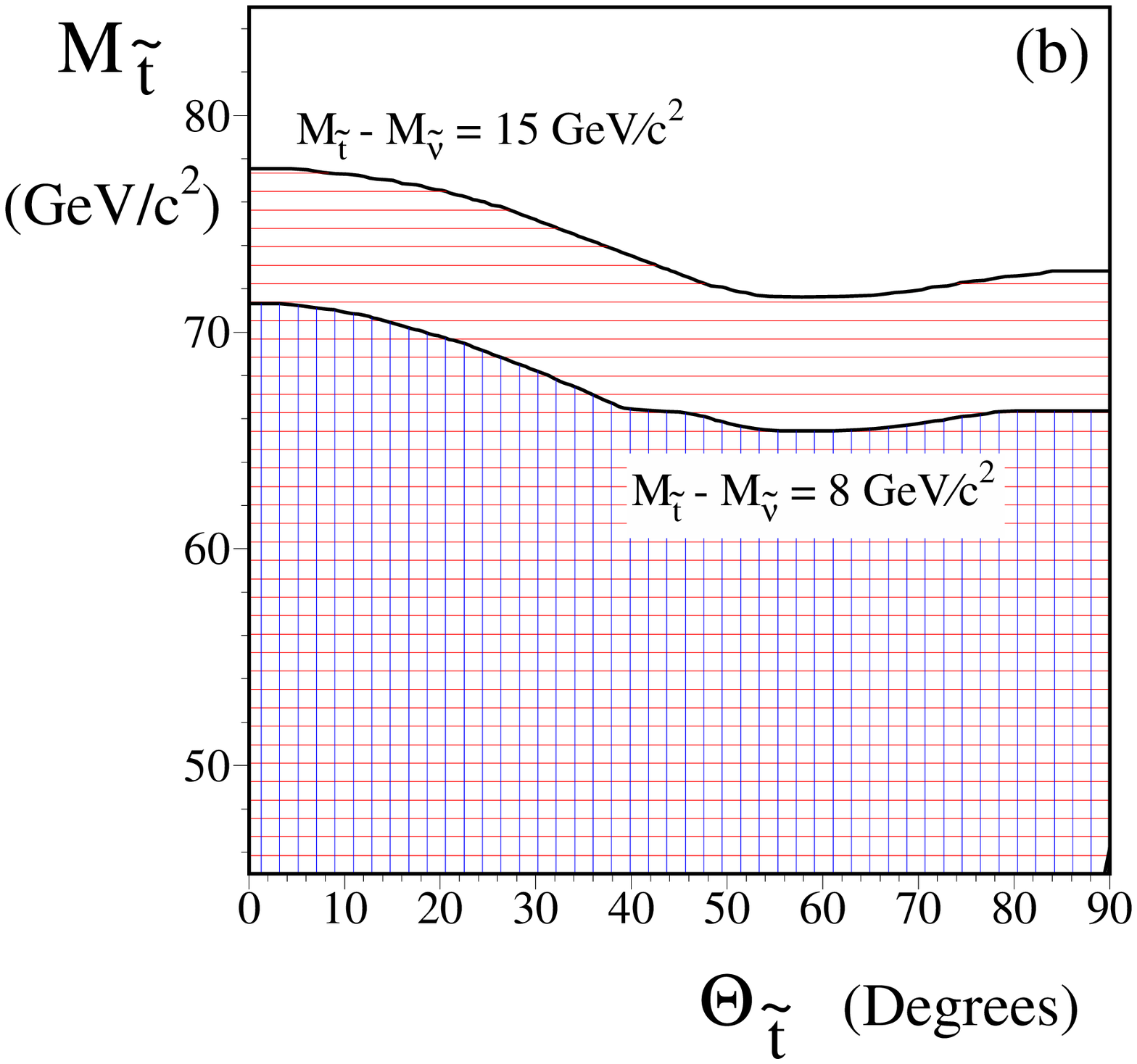,width=0.46\textwidth}
\end{tabular}
\begin{tabular}{cc}
\epsfig{file=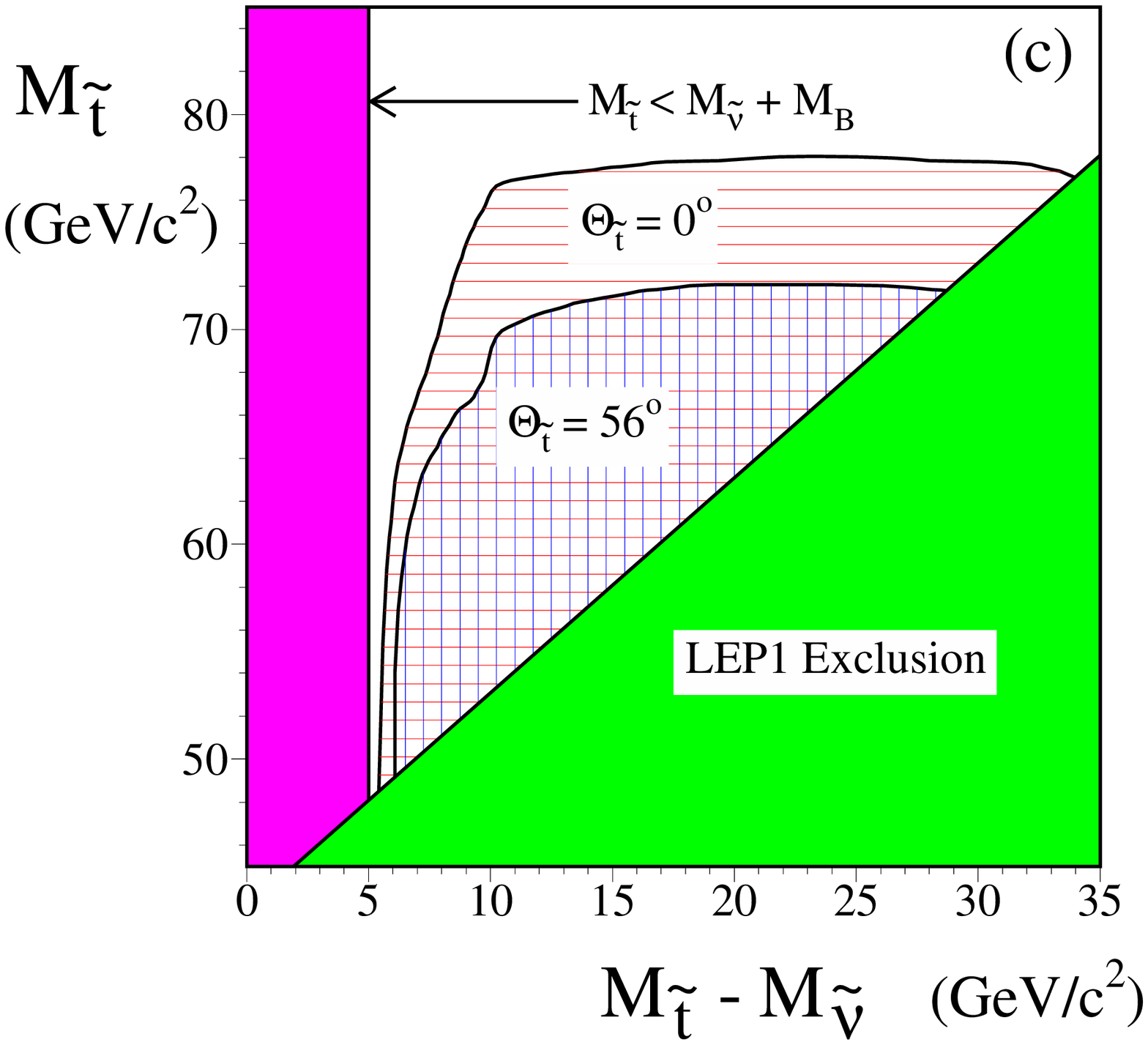,width=0.47\textwidth}
\epsfig{file=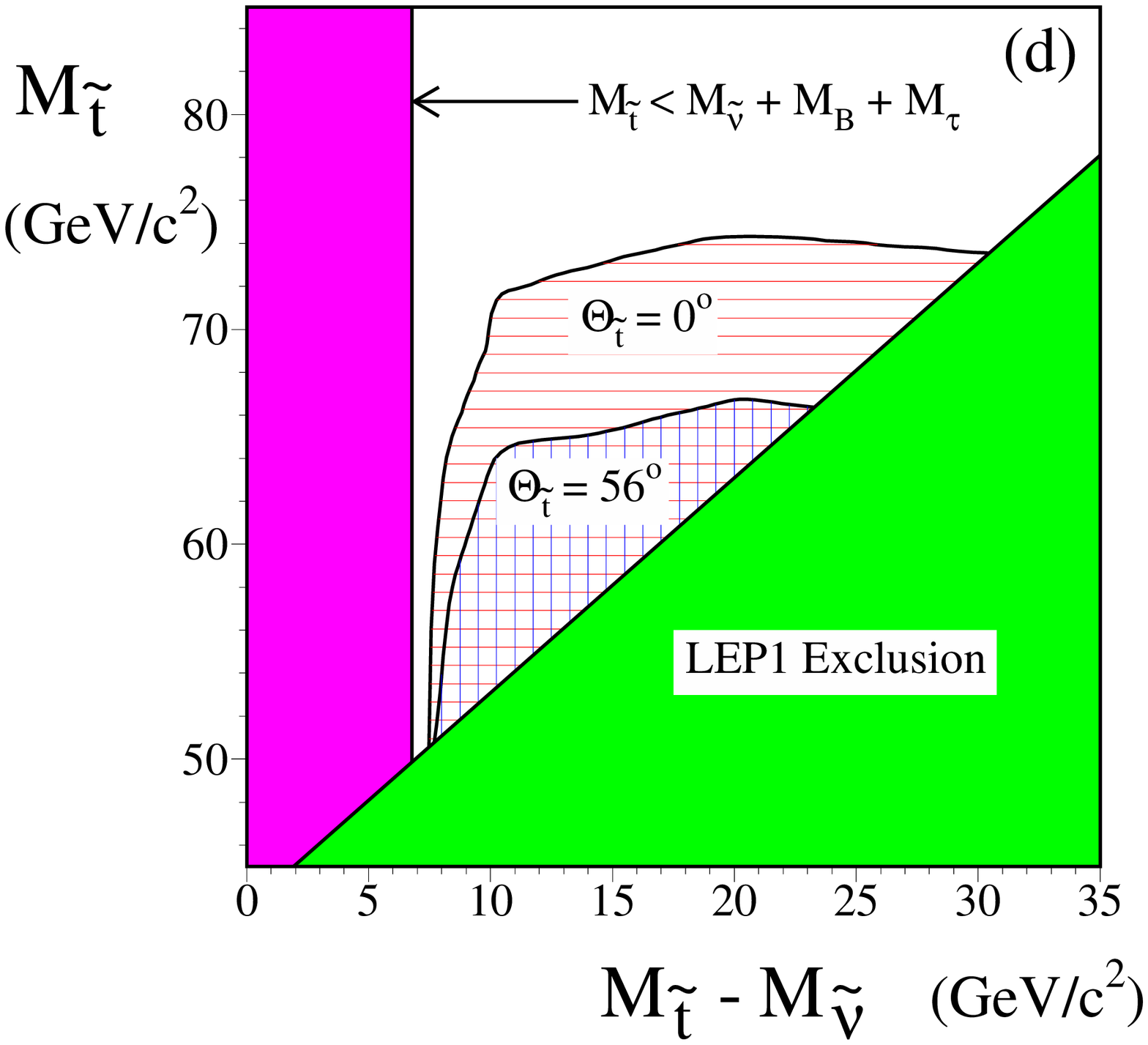,width=0.47\textwidth}
\end{tabular}
\end{center}
\caption{\rm Excluded regions assuming  
$\stop \rightarrow \mathrm{b}\ell\tilde{\nu}$.
(a) Excluded region in the $m_{\tilde{\nu}}$ vs 
$m_{\stop}$ plane. (b) Excluded region in 
the $m_{\stop}$ vs 
$\mathrm \thetamix$ plane. (c) Excluded region in the
$m_{\stop}$ vs $\deltm$ plane. In (a), (b) and (c)
equal branching fractions for the $\stop$ decay to $\mathrm{e}$,
$\mu$ or $\tau$ are assumed. 
(d) Excluded region in the $m_{\stop}$ vs $\deltm$ plane, assuming a
branching ratio of 100\% for the $\stop$ decay to $\tau$. 
In (a), (c), and (d), excluded regions are given for
$0^{\circ}$, corresponding to the maximum $\stop$-Z coupling, and for
$56^{\circ}$, corresponding to the minimum $\stop$-Z coupling. Also 
shown in (a), (c), and (d) is the excluded region from LEP1, obtained from
the measurement of the Z lineshape.
\label{stblv}}
\end{figure}

One event is selected by the $\mathrm \stop \rightarrow c\neu$,
$\mathrm \sbot \rightarrow b\neu$ selection, while
no events are selected by the 
$\mathrm \stop \rightarrow b\ell\widetilde{\nu}$ selection.
The candidate event is selected at $\rts$ = 161~GeV; its kinematic 
properties suggest the process 
$\mathrm{e}^{+} \mathrm{e}^{-} \rightarrow \zz \rightarrow 
\nu \bar{\nu} \tau^{+}\tau^{-}$ as a Standard Model interpretation.
Since only a single event is selected, 
it is appropriate to set lower limits on the 
masses of the $\stop$ and $\sbot$. 
Figures~\ref{stchi}a, \ref{stchi}b, and \ref{stchi}c give the 95\% C.L.
excluded regions for the channel $\mathrm \stop \rightarrow c\neu$.
For this channel, the $\thetamix$-independent
lower limit on $m_{\stop}$ is 67~$\gev$, assuming a mass difference between 
the $\stop$ and the $\neu$ of at least 10~$\gev$.
Figures~\ref{stblv}a, \ref{stblv}b, and \ref{stblv}c give excluded
regions for the $\mathrm \stop \rightarrow b\ell\widetilde{\nu}$ channel,
assuming equal branching ratios for the $\stop$ decay 
to $\mathrm{e}$, $\mu$, $\tau$. In this case, 
the $\thetamix$-independent lower limit on $m_{\stop}$ is 70~$\gev$, 
assuming a mass difference between the $\stop$ and the $\snu$ of at 
least 10~$\gev$. 

Figure~\ref{stblv}d gives the excluded region in the ($\deltm,m_{\stop}$) plane
for the $\mathrm \stop \rightarrow b\ell\widetilde{\nu}$ channel, assuming
a branching ratio of 100\% for the $\stop$ decay to  $\tau$. A 
$\thetamix$-independent lower limit of 64~$\gev$ is set on $m_{\stop}$
in this case, again assuming a mass difference between the $\stop$ 
and the $\snu$ of at least 10~$\gev$.

\begin{figure}[p]
\begin{center}
\begin{tabular}{cc}
\epsfig{file=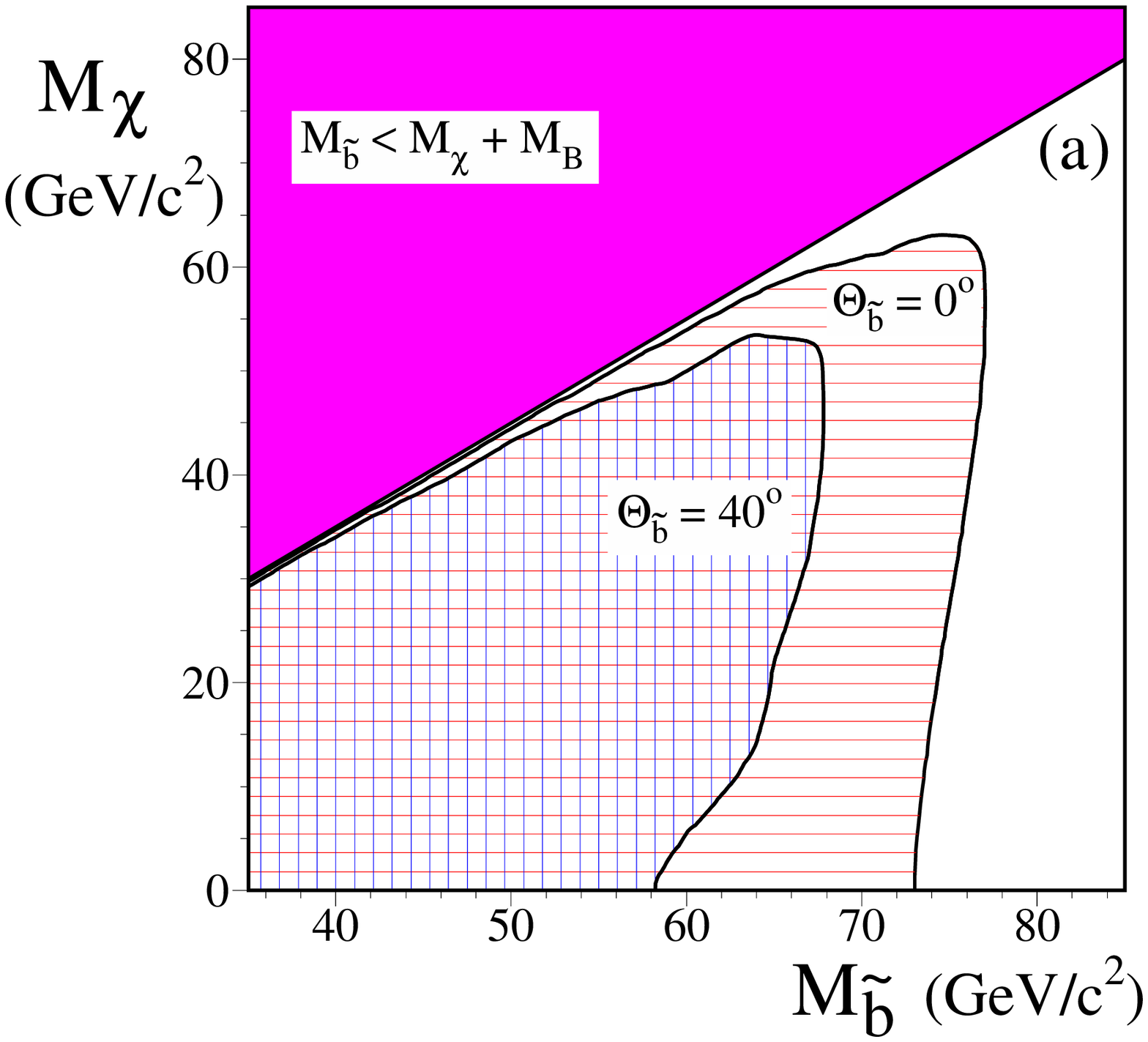,width=0.46\textwidth} &
\epsfig{file=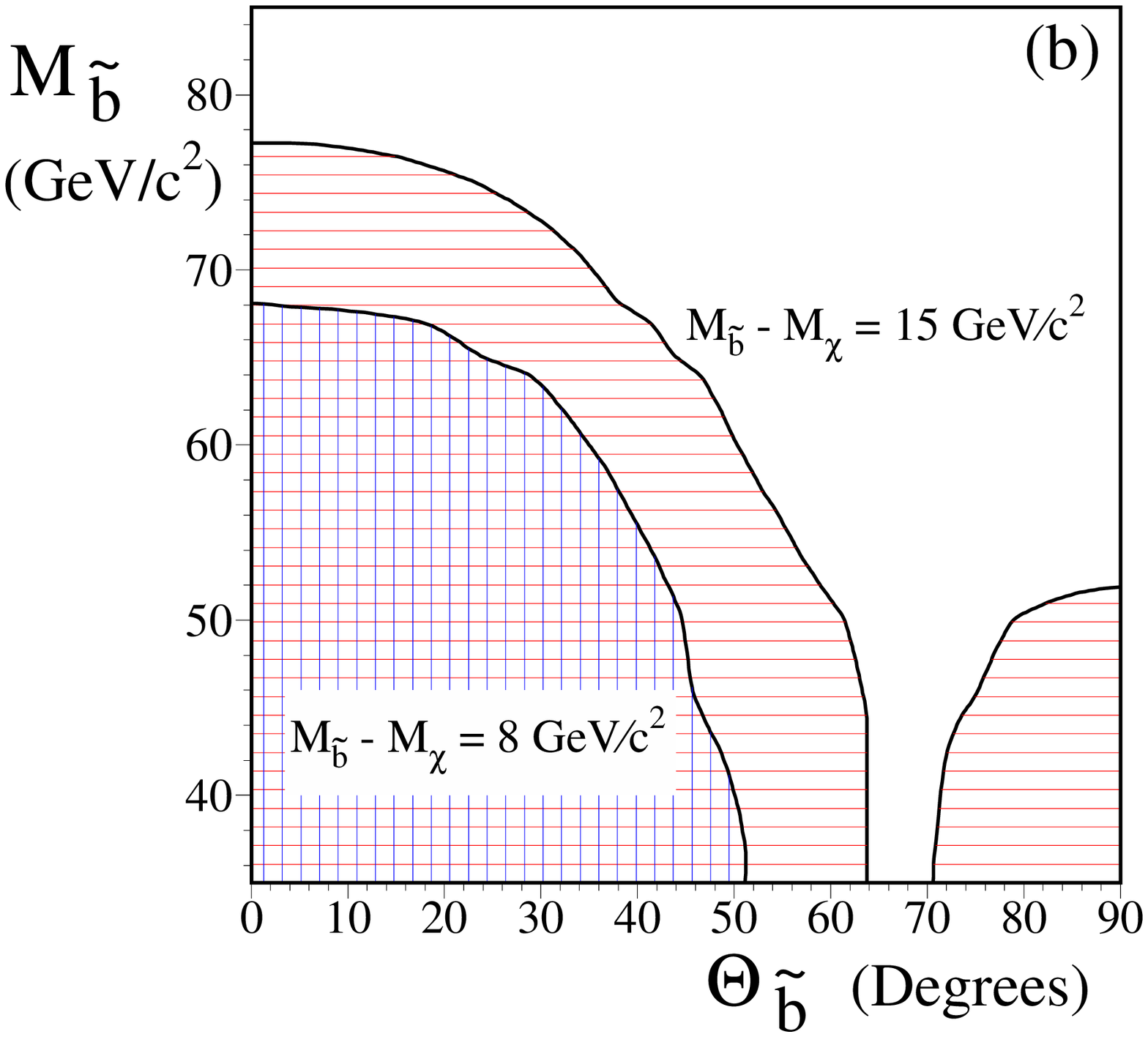,width=0.46\textwidth}
\end{tabular}
\epsfig{file=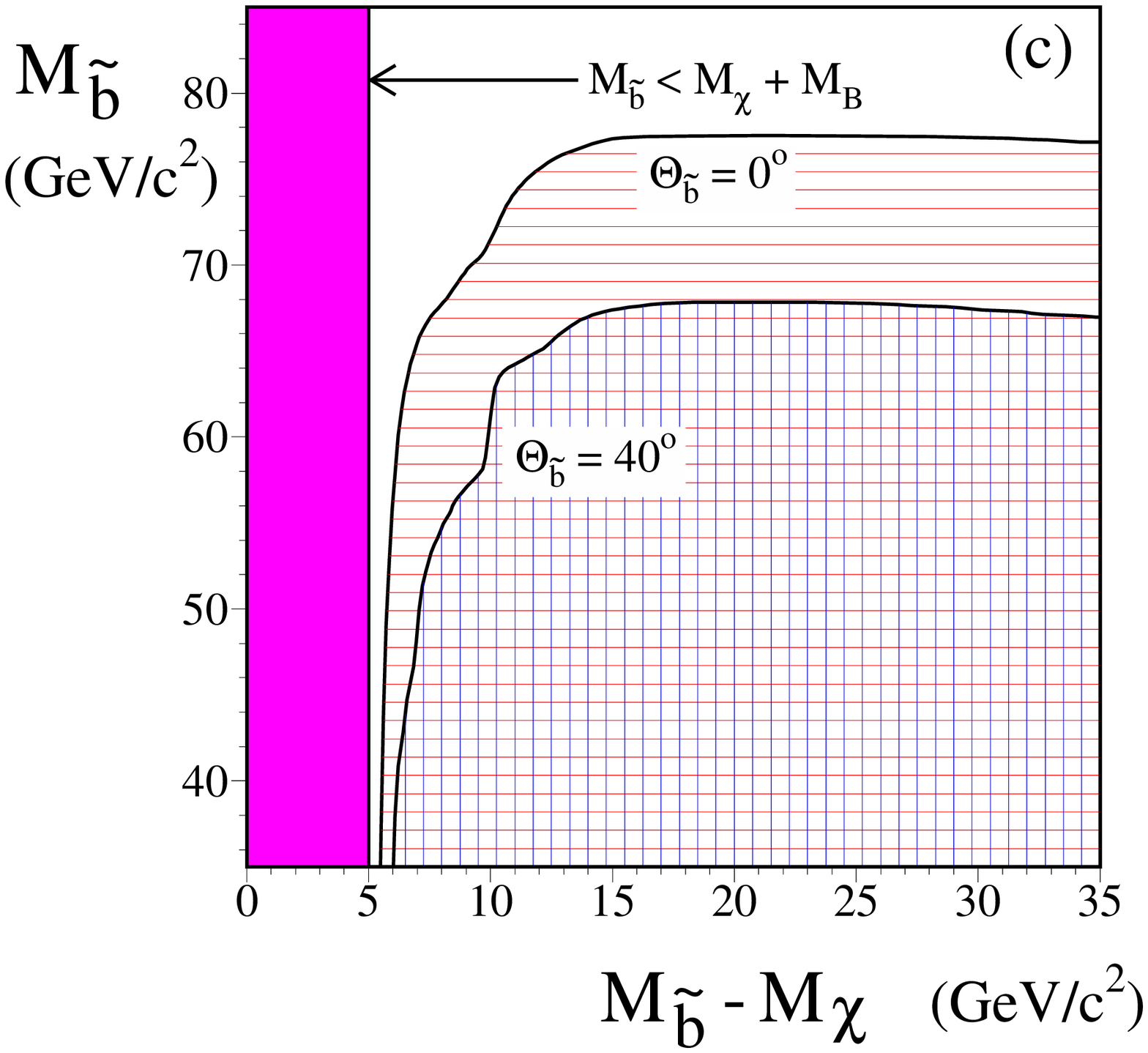,width=0.46\textwidth}
\end{center}
\caption{\rm Excluded regions assuming $\sbot \rightarrow
\mathrm{b}\neu$. (a) Excluded region in the $m_{\neu}$ vs 
$m_{\sbot}$ plane. 
(b) Excluded region in the $m_{\sbot}$ vs 
$\mathrm \thetab$ plane. (c) Excluded region in the 
$m_{\sbot}$ vs $\deltm$ plane.  
In (a) and (c), excluded regions are given for
$0^{\circ}$, corresponding to the maximum $\sbot$-Z coupling, and for
$40^{\circ}$.
\label{sbotchi}}
\end{figure}

Figures~\ref{sbotchi}a, \ref{sbotchi}b and  \ref{sbotchi}c
give the excluded regions for the
$\sbot$ decay $\mathrm \sbot \rightarrow b\neu$. 
A lower limit of 73~$\gev$ is set on $m_{\sbot}$ , assuming that
$\thetab$ is $0^{\circ}$ and that the mass difference between the 
$\sbot$ and the $\neu$ is at least 10~$\gev$.
Figure~\ref{sbotchi}b shows that 
$\thetab$-independent $m_{\sbot}$ limits are not set. 
When decoupling from the Z occurs, sbottoms can only be produced
through photon exchange and the cross section for the $\sbot$ 
(charge $-1/3$) is four times lower than the cross section for the $\stop$
(charge $+2/3$).  

%
%
\section{Conclusions}
Searches have been performed for scalar top quarks at 
$\rts$ = 130--172~GeV. 
A single candidate event, selected at 
$\rts$ = 161~GeV, is observed in the
$\mathrm \stop \rightarrow c \neu$ channel while no events are
observed in the $\mathrm \stop \rightarrow b \ell \tilde{\nu}$ channel.
This is consistent with the background expectations of 2.3 events for the
$\mathrm \stop \rightarrow c \chi$ channel and 1.0 events for the
$\mathrm \stop \rightarrow b \ell \tilde{\nu}$ channel.

\par
A 95 \% C.L. limit of $m_{\stop}$ $>$ 67~$\gev$ is obtained for the 
$\mathrm \stop \rightarrow c\neu$ channel, independent of the
mixing angle and valid for a mass difference between the $\stop$
and the $\neu$ larger than 10~$\gev$. For the 
$\mathrm \stop \rightarrow b\ell\widetilde{\nu}$ channel, the 
$\thetamix$-independent limit is \mbox{$m_{\stop}$ $>$ 70~$\gev$} if the mass
difference between the $\stop$ and the $\snu$ is greater than
10~$\gev$ and if the branching ratios are equal for the 
$\stop$ decays to $\mathrm{e}$, $\mu$, and $\tau$.

\par
A limit is also obtained for the $\sbot$ decaying as 
$\mathrm \sbot \rightarrow b\neu$. The limit is $m_{\sbot}$ $>$ 
73~$\gev$ for the supersymmetric partner of the left-handed state of
the bottom quark if the mass difference between
the $\sbot$ and the $\neu$ is greater than 10~$\gev$. 
%
%
\section{Acknowledgements}
We wish to congratulate our colleagues from the accelerator
divisions for the successful operation of LEP above the $\mathrm{W^{+}W^{-}}$ 
threshold. We would also like to express our gratitude to the engineers
and support people at our home institutes without whom this work
would not have been possible. Those of us from non-member states
wish to thank CERN for its hospitality and support.

\newpage

\newpage

\newpage

\begin{thebibliography} {99}


\bibitem{SUSY}
H.P.~Nilles, Phys.~Rep.~{\bf C 110} (1984) 1;\\
H.E.~Haber and G.L.~Kane, Phys.~Rep.~{\bf C 117} (1985) 75;\\
R.~Barbieri, Riv.~Nuovo Cimento {\bf 11}, (1988) 1.

\bibitem{Hikasa}
K. Hikasa and M. Kobayashi,
Phys. Rev. {\bf D 36} (1987) 724.

\bibitem{Drees}
M. Drees and K. Hikasa, Phys. Lett. 
{\bf B 252} (1990) 127.

\bibitem{ALEPH_stop}
ALEPH Collaboration, Phys. Lett. {\bf B 373} (1996) 246.

\bibitem{D0}
D0 Collaboration, Phys. Rev. Letters {\bf 77} (1996) 2222.


\bibitem{OPAL}
OPAL Collaboration, ``{\it Search for Scalar Top and Scalar Bottom Quarks at \mbox{$\rts$ = 170 -- 172~GeV} in $\mathrm{e}^{+} \mathrm{e}^{-}$ Collisions}'', CERN-PPE 97-046. To be published in \mbox{Z. Phys. {\bf C}.}


\bibitem{Alnim}
ALEPH Collaboration, Nucl. Instrum. and Methods {\bf A 294 } 
(1990) 121.

\bibitem{Alperf}
ALEPH Collaboration, Nucl. Instrum. and Methods {\bf A 360}
(1995) 481.

\bibitem{LEP2phys}
W. Beenakker, R. Hopker, M. Spira and P.M. Zerwas,
Phys. Lett. {\bf B 349} (1995) 463.\\
{\it Physics at LEP2}, CERN 96-01 (1996), Eds G.~Altarelli, 
T.~Sj$\mathrm{\ddot{o}}$strand and F.~Zwirner, Vol.~2.


\bibitem{JETSET}
T.~Sj$\mathrm{\ddot{o}}$strand, Comput. Phys. Commun. {\bf 82} (1994) 74.

\bibitem{Peterson}
C.~Peterson, D.~Schlatter, I.~Schmitt and P.M.~Zerwas, Phys. Rev. {\bf D 27}
(1983) 105.

\bibitem{frag}
ALEPH Collaboration, ``{\it Studies of Quantum Chromodynamics with the ALEPH
\mbox{Detector}}'', CERN-PPE 96-186. To be published in Physics Reports.



\bibitem{N95}
The ALEPH Collaboration, Phys. Lett. {\bf B 384} (1996) 427.\\



\bibitem{syse}
R.D. Cousins and V.L. Highland, Nucl. Instrum. and Methods {\bf A320} (1992) 331. 

\end{thebibliography}
\end{document}